\documentclass[superscriptaddress,nofootinbib,twocolumn,longbibliography]{revtex4-2}
\usepackage{graphicx}
\usepackage{dcolumn}
\usepackage{bm}
\usepackage{mathrsfs}
\usepackage{multirow}
\usepackage{amsmath}
\usepackage{amsfonts}
\usepackage[makeroom]{cancel}
\usepackage{soul}

\usepackage[colorlinks,
linkcolor={red!60!black!},
anchorcolor={green!80!black!},
urlcolor={blue!80!black!},
citecolor={blue!50!black!}]{hyperref}
\usepackage[table]{xcolor}
\usepackage{tipa}
\usepackage{physics}
\usepackage{caption}
\usepackage{subcaption}
\usepackage{bbold}
\usepackage{tikz}
\usepackage{sidecap}
\usepackage{placeins}

\def\be#1\ee{\begin{align}#1\end{align}}

\def\ba{\begin{eqnarray}}
\def\ea{\end{eqnarray}}
\def\nn{\nonumber}
\def\q{}

\begin{document}

\title{De Sitter horizon entropy from a simplicial Lorentzian path integral
}

\author{Bianca Dittrich}
\email{bdittrich@pitp.ca}
\affiliation{Perimeter Institute, 31 Caroline Street North, Waterloo, ON, N2L 2Y5, Canada}
\affiliation{Theoretical Sciences Visiting Program, Okinawa Institute of Science and Technology Graduate University, Onna, 904-0495, Japan}
\author{Ted Jacobson}
\email{jacobson@umd.edu}
\affiliation{Maryland Center for Fundamental Physics, University of Maryland, College Park, MD 20742, USA}
\affiliation{Perimeter Institute, 31 Caroline Street North, Waterloo, ON, N2L 2Y5, Canada}
\author{Jos\'e Padua-Arg\"uelles}
\email{jpaduaarguelles@pitp.ca}
\affiliation{Perimeter Institute, 31 Caroline Street North, Waterloo, ON, N2L 2Y5, Canada}
\affiliation{Department of Physics and  Astronomy, University of Waterloo, 200 University Avenue West, Waterloo, ON, N2L 3G1, Canada}

\begin{abstract}

The dimension of the Hilbert space of a quantum 
gravitational system can be written formally
as a path integral partition function 
over Lorentzian metrics. We implement this
in a 2+1 dimensional simplicial 
minisuperspace model in which the system is a spatial topological disc, and recover by contour deformation 
through a Euclidean saddle the entropy of the
de Sitter static patch, up to discretization 
artifacts. The model illustrates the importance
of integration over both positive and negative lapse
to enforce the gravitational constraints, 
and of restriction to complex metrics for 
which the fluctuation integrals would converge.
Although a strictly Lorentzian path integral is oscillatory,
an exponentially large partition function
results from unavoidable imaginary contributions 
to the action. 
These arise from analytic continuation
of the simplicial (Regge) action 
for configurations with
codimension-2 simplices where the metric fails to be 
Lorentzian. In particular, the dominant contribution 
comes from configurations with contractible 
closed timelike curves that encircle the boundary 
of the disc, 
in close correspondence with 
recent continuum results. 
\end{abstract}

\maketitle

\section{Introduction \label{sec:intro}}

Although nearly 50 years have passed 
since it was first introduced by Gibbons and Hawking, 
the path integral representation of the 
quantum gravitational partition function remains puzzling. 
The general consensus is that its saddle point approximation 
captures real physics of the quantum theory such as the 
Bekenstein-Hawking entropy
of black hole and de Sitter horizons; but the reasons for that success,
and even the very definition of the path integral, have not been 
fully understood. 
Of course this path integral is at best an
effective description of some underlying UV complete theory.
And it is probably fair to say that a good portion of the 
obscurity is due to this fact. However, the effective description
is presumably embedded in the more complete one in 
a rich and complicated way, and remains approximately valid in suitable regimes.
Lacking the more complete theory, the effective theory is 
thus one of our best guides.

The most pressing question is why the path integral apparently
captures the correct horizon entropy $A/4\hbar G$ 
without putting its finger on the
states the entropy is counting. Part of the answer is that 
$G$ in that formula is a phenomenological constant, which 
depends on the underlying UV complete theory.
The path integral provides a link between
the macroscopic gravitational dynamics and the statistics of the underlying 
unknown microstates, in the semiclassical
approximation, where the geometrical contribution to the partition
function is dominant. 

But the phenomenological nature of $G$ is not the whole story. 
The Gibbons-Hawking calculation relies also on the assumption that
it is correct to approximate the partition function 
$Z$ by a Euclidean (signature) saddle point, despite the fact that the 
paths in the path integral are not Euclidean geometries. 
The Euclidean action is unbounded below,
due to the conformal mode, so the path integral 
would not converge were it to be taken over
Euclidean geometries. 
In fact, since 
the partition function is a
trace in the physical Hilbert space, its path integral
representation should involve only paths satisfying the
initial value constraints, \textit{i.e.}, after gauge fixing, 
paths in the reduced phase space.
The constraints eliminate the
conformal mode, so evidently the Euclidean path integral is 
not equivalent to the reduced phase space one \cite{Hajicek:1984nb,Hartle:2020glw,Schleich:1987fm,Mazur:1989by}.

A justification of the saddle point calculation 
must therefore begin with a ``real time'' path integral representation that indeed
imposes the constraints, and it must then be shown from that starting
point that the contour of integration may be deformed so as to 
pass through a Euclidean saddle that dominates the integral. 
This appears at first impossible, because real time path integrals 
have oscillating integrands and so cannot produce the $\exp(A/4\hbar G)$ behavior required to recover the expected entropy. Indeed, Picard-Lefschetz theory establishes that a saddle point can contribute to the path integral only if the saddle point can be approached from the original contour along a steepest {\it descent} contour of the real part of the integrand's exponent. If the exponent is pure imaginary on the original contour, this implies that its real part at the contributing saddle must be {\it negative}.

However, the partition function being a trace, its path integral representation is a sum over path geometries with closed timelike curves (CTC's). As we argue in section \ref{sec:CF}, if the gravitational system is bounded by a horizon, the time slices should share a common boundary at the horizon, in which case
the path geometries have closed timelike curves
that are contractible to points on the horizon.
At such a point no Lorentz signature metric exists. 
We shall call this a ``CTC singularity''. It is similar to a conical 
singularity, in that it can be formed by gluing two edges of an otherwise 
flat portion of Minkowski spacetime, and for this reason Marolf
called it a ``Lorentzian conical singularity'' \cite{Marolf:2022ybi}. But, unlike 
a Euclidean conical singularity, there is no way to smooth it with a
high curvature tip, so it is more radically singular. In fact, it is
so radically singular that the gravitational action associated with
such a geometry acquires an imaginary part. Precisely because of this
imaginary part, the integrand $\exp(\imath S/\hbar)$ can develop an exponentially
enhanced real part, which can produce the expected entropy after all.

To verify this route to the horizon entropy, 
one needs to justify the assignment of the
imaginary part of the action and the contour of integration,
and to show that the contour can be deformed to a dominating Euclidean saddle. 
In this paper we shall study this in the context of
de Sitter horizon entropy or, what is the same, 
the calculation of the trace of the identity operator on 
the Hilbert space of a ball of space in general relativity 
with a positive cosmological constant. We implement the 
analysis in the extremely simplified, 
yet still quite instructive, setting of simplicial $2+1$ dimensional
spacetimes constructed from four tetrahedra with just two independent
variable edge lengths. The advantage is that we deal only with ordinary 
integrals, with no room for uncertainty about infinite dimensional
path integrals and ultraviolet divergences, and yet this simplicial
minisuperspace system seems able to capture the key physics.
Although of course this discrete minisuperspace does not 
contain the microscopic degrees of freedom responsible for the horizon 
entropy, nor even their effective field theory descendants, it does
contain the CTC singularity which is the topological feature 
responsible for the ``imprint'' of the entropy on the saddle action.
Since, as already alluded to above, the measured value of the gravitational
constant reflects the underlying degrees of freedom responsible
for the entropy~\cite{Susskind:1994sm,Jacobson:1994iw}, a simple calculation of this nature can evidently reflect the value of the entropy.

Despite its discreteness, our approach to the problem was motivated by,
and is technically quite related to, the recent work of Marolf \cite{Marolf:2022ybi}, 
who approached it in the continuum setting, examining the 
contribution of black hole entropy to the 
thermal partition function at a given temperature. 
In particular, Marolf's analysis uses
real time contours, the imaginary contribution to the action
coming from the CTC singularity, and the organization of the 
path integral into first an integral at fixed horizon area,
followed by an integral over the area, all of which are key elements 
in our approach as well.

\section{Computational framework and key results}\label{sec:CF}

The dimension of a Hilbert space ${\cal H}$ is equal to $Z:=\Tr I_{\cal H}$,
where $I_{\cal H}$ is the identity operator on that space. For a
quantum system arising from a classical phase space with canonical
coordinates $(q,p)$, one can express this trace as a path integral, 
by inserting alternating complete sets of $q$ and $p$ eigenstates in the
usual way, passing to a limit of continuous paths $(q(t),p(t))$,
and identifying the initial and final states, resulting in\footnote{From here
on we choose units with $\hbar =1$.}
\begin{equation}
Z = \int {\cal D}p{\cal D} q\, \exp({\imath\oint p \dot q dt})\, .
\end{equation}
The time period here is irrelevant, since $\oint p \dot q dt = \oint p dq$
does not depend on it.\footnote{When
one computes not the trace of the identity but rather the evolution operator,
what appears in the exponent is $\imath$ times the action.}
When applied to a theory with gauge symmetry this construction must 
employ the {\it reduced} phase space space, for which the constraints
and gauge fixing conditions have been imposed. The constraints $C_i$ can be
imposed with delta functions, which can be expressed as Fourier integrals over 
Lagrange multipliers $\lambda_i$, such that
the  path integral 
integrand becomes $\exp[\imath \oint (p\dot q - \sum_i \lambda_i C_i)]$
times a gauge-fixing determinant that we shall
regard as having been absorbed into the measure.
When this is done for general relativity \cite{Faddeev:1973zb}, the momenta appear quadratically,
so one can integrate them out
and, up to possible boundary terms, the 
integrand then becomes $\exp(\imath S^\text{L})$, where $S^\text{L}$ is the Lorentzian 
action. Thus now, because of the Lagrange multiplier integrals,
one is necessarily integrating over all proper time periods.
Moreover, since each multiplier integral is over the whole real line,
one must include both positive and negative periods. It
might therefore appear
that one is actually 
computing twice the real part of 
the path integral over only positive time periods; however, as discussed 
in \S\ref{ssec:saddle_analysis},
there is a branch point with an essential singularity at vanishing lapse, around which the lapse integration contour must navigate.

This construction was reviewed recently in \cite{Banihashemi:2022jys}, where it was pointed out
that when computing not $\Tr I$ but rather the thermal partition function
$\Tr\exp(-\beta H)$, there is a mismatch between the real exponent involving the 
ADM or Brown-York Hamiltonian at the outer boundary of the system, and the imaginary
exponent indicated above. However, for a system 
like the de Sitter static patch, with no outer boundary, no such mismatch exists, and 
$H=0$ so the thermal partition function becomes 
$\Tr I$, the dimension of the Hilbert space.\footnote{Another setting in 
which no mismatch exists is when computing the density of states, 
rather than the thermal partition function \cite{Brown:1992bq}.}
The role of the ``horizon'', however, requires more discussion.

We focus here on the case of a horizon like that of a static patch
of de Sitter space. We presume that, despite the fact that 
due to the diffeomorphism constraints
the full quantum
gravity Hilbert space is not spatially factorizable, 
it is meaningful to consider the Hilbert space of degrees of freedom
of a gravitational system in a region of space 
bounded by what will wind up being a slice of a horizon
in a saddle point approximation. It is arguable (but not uncontroversial)
that this is the sort of system to which Gibbons and Hawking's seminal 
black hole thermal partition
function refers. Indeed the saddle there is foliated by hypersurfaces whose geometry 
coincides with the spatial hypersurfaces of the Lorentzian black hole
{\it outside} of and terminating on the bifurcation surface of the horizon. If the path integral is to access a saddle of this topology, the time foliation must consist of spatial slices that 
all coincide at a codimension-2 boundary of the system, which in the saddle
configuration becomes identified with the horizon of the Lorentzian black hole.

For the case of a region
with no {\it outer} boundary, in $D$ dimensional spacetime, 
each spatial slice is a $(D-1)$-ball, so another way to describe what is being counted 
is the dimension of the Hilbert space of states of such a ball of space \cite{Jacobson:2022jir}.
In this paper we concentrate on the case $D=3$, so for concreteness let us
consider that case here. Each spatial slice is then a 2-ball, \textit{i.e.}, a disc.
If all of the discs in the foliation share the same boundary circle (1-sphere), and in the
periodic time dimension they wrap around that circle, \st{so} then the boundary circle is
encircled by contractible closed timelike curves. 
The Lorentzian metric is therefore 
not well defined at the boundary circle, which is
thus a sort of singularity. As 
mentioned above we term this a CTC singularity. 

While the existence of a saddle with the Gibbons-Hawking topology certainly motivates the restriction to path geometries containing a CTC singularity, that restriction should be justified from first principles. Although we are currently unable to provide a complete justification, we can offer three further supporting arguments. All 
are concerned with the nature of the boundary of the system whose states our partition function is counting. 

No boundary conditions are imposed at the boundary of our system (on one time slice); rather, that boundary 
is supposed to correspond to the corner of the causal horizon of the purview of some observer. If, in the path integral, each time slice were to have a distinct boundary, the boundary of a path geometry would have codimension-one, and the 
dynamical system would be ill-defined without the imposition of boundary conditions and addition of appropriate boundary terms to the action. To avoid these unwanted ingredients we can
require that all the time slices share the same codimension-two boundary. 
This also neatly coincides, in a spacetime picture, 
with the time slices being Cauchy surfaces of the causal domain of the observer, all of which meet at the corner. 
Moreover, because the gauge constraints at the boundary involve the degrees of freedom on both sides of the boundary, they are not a property of our subsystem by itself. In order to not impose them in the path integral, 
we should not integrate over the lapse and shift--- \textit{i.e.}, the Lagrange multipliers involved in imposing the constraints ---at the location of the codimension-two boundary, and this can be achieved in a diffeomorphism-invariant manner by setting the lapse and shift to zero at the boundary. The resulting geometries then have no time flow at the boundary,
which implies that the periodic time flow of each path geometry leaves the boundary fixed, hence the boundary lies at 
a CTC singularity.

Topologically, the manifold foliated by a cycle of discs all sharing the same
boundary is a 3-sphere.
To visualize this it is helpful to begin with the lower dimensional
case $D=2$, so the spatial slices are 1-balls, \textit{i.e.}, line intervals,
whose boundary consists of a 0-sphere, \textit{i.e.}, a pair of points. A cycle of 
such intervals, encircling the boundary points, forms a 2-sphere,
as depicted in FIGURE~\ref{fig:2D_continuum}. One can visualize the $D=3$ case 
by ``decompactification'': Replace the 2-disc slices by half-planes, which
become 2-discs when a point at infinity is added. The boundary circle 
of the 2-discs then becomes the rectilinear infinite edge of the half-plane,
say the $z$ axis in $\mathbb R^3$ (see FIG.\ \ref{fig:3D_continuum}).
The half-planes all coincide on the $z$ axis, and they wrap around it, filling
all of $\mathbb R^3$, which together with a point at infinity forms the 3-sphere.

Thus, to compute $Z$ for states of a 2-disc, we should carry out
the path integral with integrand $\exp(\imath S^\text{L})$ over metrics on 
$S^3$ that have a CTC singularity on a spacelike circle, including
both signs of the time flow direction.

Of course this does not quite make sense, however, because 
the Lorentzian action is ill-defined on a spacetime with a CTC singularity.
In the continuum case, Marolf \cite{Marolf:2022ybi} used the Lorentzian Gauss-Bonnet
theorem to motivate a definition of this action. In our simplicial setting,
a definition is provided by analytic continuation of the Regge action.
In either case, it is at this point only a somewhat well motivated
definition. If this approach to evaluating horizon entropy is 
to be physically meaningful, however, it should ultimately find a better justification.

Simplicial minisuperspace constructions of the Euclidean path integral have been considered by Hartle in \cite{Hartle1,Hartle2,Hartle3}. Here we are interested in a simplicial Lorentzian minisuperspace path integral \cite{Asante:2021phx}. 
The simplicial spacetimes we integrate over
are constructed with  four tetrahedra and two independent
edge lengths, so there are just two variables of integration,
which can be thought of as the perimeter of the spatial disc (an equilateral
triangle for us), and the time period in the center of the disc.
The action is the analytically continued Regge action \cite{Regge:1961px} (see \S\ref{ssec:Regge_intro}), 
including a positive cosmological constant term.
This allows us to construct a simplicial version of the partition function
described above, through a (mostly) Lorentzian path integral.

As discussed above, the time period can be interpreted as Lagrange multiplier.  As in the continuum we integrate this time period over all real values. Curiously, in the simplicial setting, only a small range of the integration domain for the time period describes configurations with CTC singularities. Another small range describes configurations with a different kind of light cone irregularities along the edges, which are initial or final singularities, that is, end points for future or past directed trajectories.
In addition, we have an unbounded range describing big bang to big crunch cosmologies, which are light cone irregular 
 only
at the big bang and the big crunch points. 
It is thus impossible, with the simplicial complex we have adopted, to emulate
the continuum calculation with respect to both the range of the time period and
the chronotopology. We choose to integrate over the full time period, because
we consider the imposition of the constraints to be the more fundamental
ingredient, while the inclusion of the other chronotopological configurations may be 
viewed as a discretization artifact. 
The would-be purely oscillatory Lorentzian path integral
receives exponentially enhanced contributions, thanks to an imaginary part of 
the action in the presence of edges where there is no regular Lorentzian light cone structure. 
This includes the configurations with CTC singularities, 
but also the initial and final edge singularities
mentioned above.

For the configurations with CTC singularities the enhancing exponent is proportional to the bounding circumference of the disc, which can be arbitrarily
large. This raises the question how can the path integral possibly
converge when it apparently receives arbitrarily large exponential 
contributions. The answer is that signs matter. Oscillations and 
overall minus signs arising from the time period integral
can cancel contributions, and in fact that is
what happens.
It turns out that above a critical disc perimeter, set by the
length scale of the cosmological constant, the integral over the 
time period vanishes {\it exactly}. 

Something similar occurs in the continuum case 
(in any spacetime dimension)
studied by Marolf \cite{Marolf:2022ybi}, in a different physical context.
He computes the thermal partition
function allowing for states containing a black hole. 
In addition to a path integral over geometries that computes
the real time evolution operator, there is an integral like 
a Laplace transform that yields the canonical
partition function. 
The Boltzmann suppression that occurs for large 
energy contributions suppresses the exponential enhancement from
large horizon areas. 
In our case, by contrast, it is the cosmological 
constant, together with the closed spatial topology, that is
responsible for cutting off the large disc contributions.

We find that the (mostly) real time contour does indeed receive 
an exponentially enhanced contribution as expected from 
the Bekenstein-Hawking entropy and the original Gibbons-Hawking 
result, which can be seen by an explicitly justified 
contour deformation that passes through a dominating 
Euclidean saddle, thus supporting the conjecture in
 \cite{Banihashemi:2022jys,Jacobson:2022jir}. 
This result depends crucially on how the contour
navigates the branch cut that arises in the
Regge action evaluated on light cone irregular configurations
(which include the CTC singularities). We make this
choice based on a convergence criterion that was first 
enunciated by Halliwell and Hartle \cite{HalliwellContours}, and has since been
discussed and generalized by others. Namely, the integrals over quantum field fluctuations (which are not explicitly included in the minisuperspace treatment, but which behind the scenes are responsible for renormalization of the parameters in the effective minisuperspace action) 
should converge mode by mode, as they do for the vacuum fluctuations in a regular state in flat spacetime. In the literature this criterion has mostly been imposed to select 
viable semiclassical saddles, but here we apply it everywhere along the
contour of integration, since it appears to be required if the minisuperspace path integral is to have any chance of providing
a decent approximation to the full theory.

Finally, we should address a point of principle concerning the UV completion of the theory. 
We integrate over all disc sizes, including arbitrarily small ones; and, 
because the integration runs over all values of the lapse, 
the original integration contour---even after deviation around the branch point at zero---comes arbitrarily close to vanishing lapse. The 
integral thus includes regimes where the spacetime
volume is arbitrarily small in Planck units. 
Since we are using the simplical form of Einstein gravity, which is presumably only a low energy effective theory,
this raises the question whether our model is physically consistent with 
its UV completion. In the three dimensional spacetime
setting studied in this paper the question is moot, since there are no
local degrees of freedom. In higher dimensions, however, the question must be faced. A key fact is that in the model the dominant contribution to the integral can, after deformation of the contour, be attributed to a semiclassical saddle that is far from the Planckian regime. To justify the computation we must therefore 
assume that, whatever the UV completion yields for the small disc and small lapse part of the integrand, the semiclassical saddle dominates over whatever contribution arises in the UV. This assumption is plausible, because there is no reason to expect a huge exponential from the UV, both since the action there is $O(1)$ in Planck units, and because whatever is the correct UV theory it must produce semiclassical dominance, since that yields the observed low energy gravitational phenomonology. 

\section{The discretization and light cone structures \label{sec:discretization}}

Our first step in computing a
simplicial partition function for the states
of a topological disc of space is kinematic: 
we construct a simple discretization of the three-dimensional universe with topology $S^3$, which incorporates configurations with
a CTC singularity on the boundary of a triangle. In the saddle point
approximation this  corresponds to
the cosmological horizon of an ``observer''.

While our restriction to three spacetime dimensions is done for simplicity of computation and visualization, we do not expect the four-dimensional case to be either significantly more involved or qualitatively different. Indeed the key features of the path integral relevant here show up in the same form for similar four-dimensional cosmological scenarios considered in \cite{Dittrich:2021gww,Asante:2021phx,Dittrich:2023rcr}. These works employed the Lorentzian simplicial path integral in order to construct a no-boundary wave function for de Sitter space. For a much earlier work using an Euclidean simplicial path integral towards the same end, see \cite{Hartle3}.

\begin{figure}[t]
    \centering
    \begin{subfigure}[c]{1\columnwidth}
        \centering
        \includegraphics[width=0.65\linewidth]{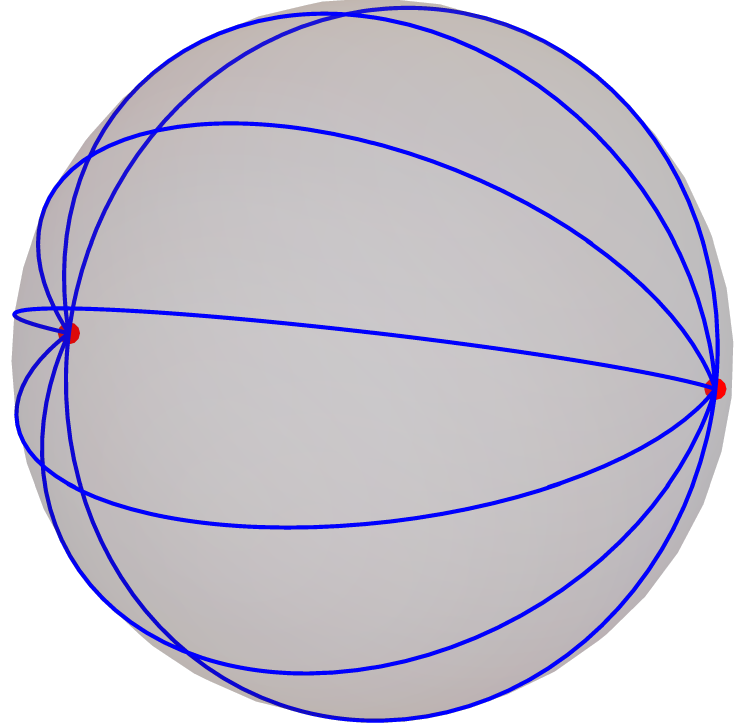}
        \caption{Foliation of a spacetime 2-sphere by 1-dimensional spatial balls (blue segments) rotated around their common boundary, a 0-sphere (red points), the \textit{horizon}. There are contractible CTCs around the horizon points, which we therefore refer to as CTC singularities.}
        \label{fig:2D_continuum}
    \end{subfigure}
    \hfill
    \begin{subfigure}[c]{1\columnwidth}
        \centering
        \includegraphics[width=\linewidth]{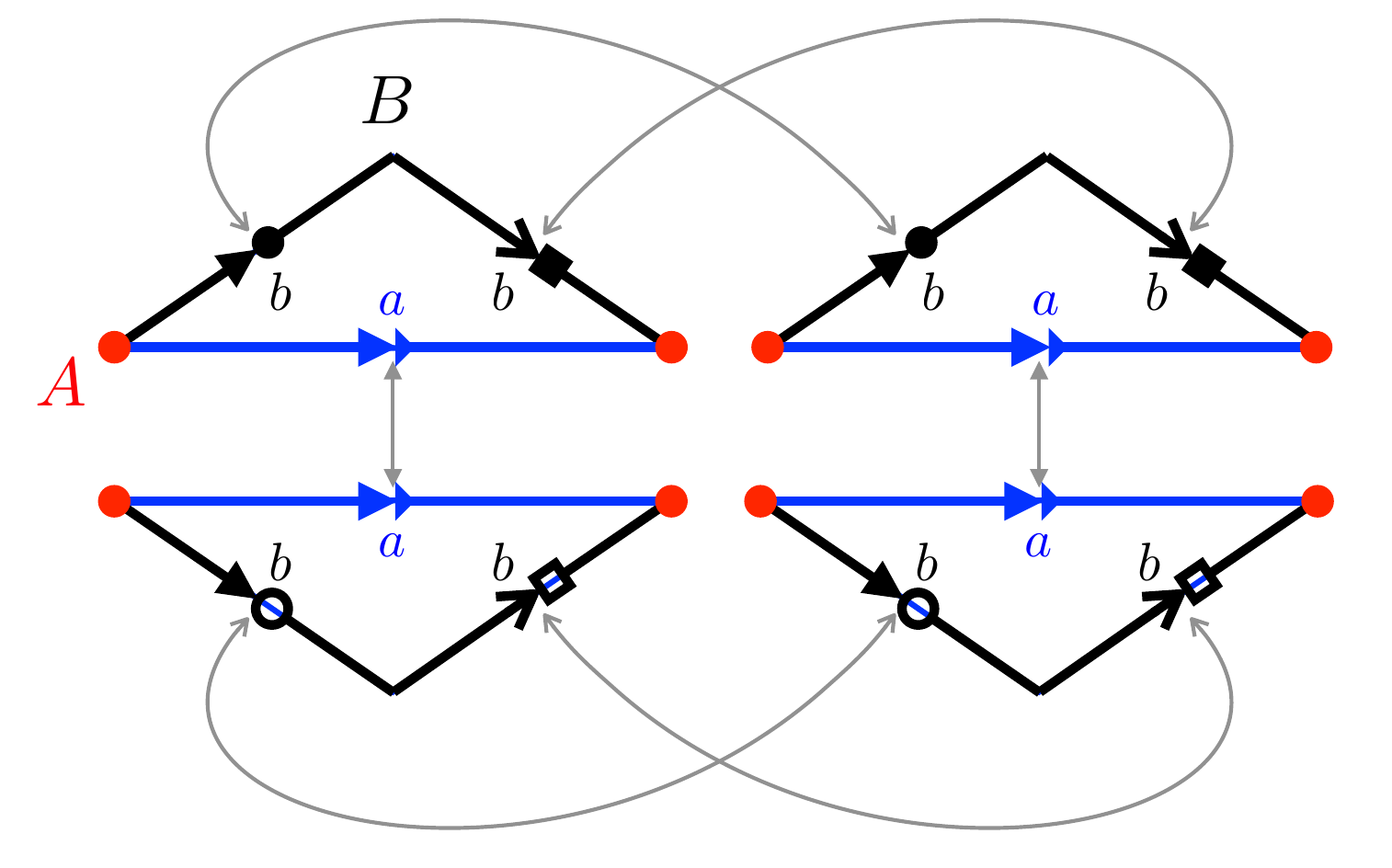}
        \caption{Triangulation of the spacetime depicted on the top figure, in which we glue four triangles along their edges as instructed by the (oriented) arrows (note that the left and right blue/base edges are only glued with each other at their two
        boundary points) —see also the external lines. The horizon is replaced by the red points and the metric by the lengths of the segments $a$ and $b$.}
        \label{fig:2D_discrete}   
    \end{subfigure}
    \caption{Discrete and 
    continuum 2-sphere spacetimes with   chronotopology analogous to that of D-dimensional configurations in the gravitational partition function that computes the dimension of the Hilbert space of a $(D-1)$-ball of space. }
    \label{fig:2D_spheres}
\end{figure}

Before taking up the three-dimensional case, let us consider the two-dimensional one as a warm-up.  A simple discretization  of the two-dimensional continuum manifold with topology $S^2$ in FIG.~\ref{fig:2D_continuum} is shown in FIG.~\ref{fig:2D_discrete}.  It consists of four Lorentzian triangles, all of which have the same geometry determined by their three edge lengths `squared'\footnote{Here and below, when we refer to geometric quantities such as length, norm, etc. \textit{squared} we mean it in the same sense in which $\mathrm d s^2=g_{\mu\nu}\mathrm dx^\mu\mathrm dx^\nu$ is a line element ``squared'', not in the sense of them being the square of a real number. Importantly, these are negative if the geometric object in question is timelike. For instance, the `length squared' of a timelike edge is negative.} $(s_a,s_b,s_b)$, and which are glued at their boundaries. Note that equating their geometry and imposing that they are isosceles constitutes a further reduction in the number of degrees of freedom, in addition to the one imposed by approximating spacetime with our triangulation. We will do the same in the three-dimensional triangulation.\footnote{An even more minimal triangulation 
of $S^2$
would consist of only the two isosceles triangles on the left in FIG.~\ref{fig:2D_discrete},
glued   at their corresponding edges. However, in that case all vertices would have an irregular causal structure in the $s_b<0$ regime (there would be zero light cones at each $B$ vertex and one at each $A$ vertex), while the $s_b>0$ regime would stay qualitatively the same as in the four triangle simplex.
An analogous situation holds for $S^3$, which can be triangulated with just 
two tetrahedra glued at their corresponding faces, but the triangulation we 
employ introduces fewer irregularities. 
It is also the case that it gives a better approximation to the 
continuum, although that is not essential for our present purposes.}

The lengths squared $s_a$ and $s_b$ must satisfy the Lorentzian triangle inequalities. Let us illustrate their derivation in the case in which the $a$-edge is spacelike, so that $s_a>0$.
Any triangle with at least one spacelike edge in two-dimensional Minkowski space can be described in a suitably aligned Minkowski coordinate system by vertices
 $\{(0,0),(0,x_2>0),(t_3>0,x_3>0)\}$. Therefore the triangle inequalities are satisfied if and only if one can find positive \textit{real numbers} $x_2$, $t_3$ and $x_3$ such that this triangle has edge lengths squared equal to our given $s_a>0$ and $s_b$. Equating $s_a$ and $s_b$ with the corresponding norms squared of the edge vectors one sees that
\begin{equation}
    x_2=\sqrt{s_a},\quad x_3=\frac{\sqrt{s_a}}2,\quad\text{and}\quad t_3=\frac12\sqrt{s_a-4s_b}\,\, .
\end{equation}
Hence the reality requirement along with the non-degeneracy requirement $t_3>0$ is equivalent to demanding $s_a>4s_b$. This inequality can be interpreted in a different way that will be useful later: we can think of $t_3$ as the height of the triangle, and correspondingly of the norm squared of $(t_3,0)$ as a \textit{height squared}, $s_h=-t_3^2$. The triangle inequality therefore states that the height squared must be negative. Note that this is trivially satisfied when the $b$-edge is timelike, because then $s_b<0$ and we have that $s_a>0$ \textit{a priori}.

With the triangles being timelike we have two light cones at each inner point of the triangles and also at each inner point of the identified edges.  
However, a fully regular Lorentzian metric on $S^2$ does not exist.  Indeed, in our triangulation some vertices must be (light cone) irregular, in the sense that these vertices carry a number of light cones that differs from two. 
There are two types of vertices: the vertices of type $A$ have adjacent edges of type $(a,b,a,b)$ and vertices of type  $B$ have adjacent edges of type $(b,b)$.

 With the $a$ edges fixed as spacelike, the spacetime geometries 
of our $S^2$ complex are classified by the nature of the $b$ edges,
which can be either spacelike, timelike, or null. Leaving aside 
the null case, which as discussed below does not contribute to the path integral, there are two cases to consider. 
If $b$ is spacelike,
the two vertices of  type $B$ are light cone regular, while those
of type $A$ do not have a light cone attached to them: all directions emanating from the $A$ vertices are spacelike
and there are contractible closed timelike curves (CTC's) encircling those vertices. 
We can therefore identify the type $A$ vertices with a horizon. 

If instead $b$ is timelike, the vertices $A$ are regular and vertices $B$ are irregular: all directions emanating  from the $B$ vertices are timelike. 
In this regime we do not have CTC's around the $A$ vertices anymore: The curves that in the regime $s_b>0$ are CTC's consist now of two parts with opposite time orientation. 

Both of the above cases place a Lorentzian metric on the 
2-sphere, and both are time orientable, 
but they have topologically different time flows and
different sorts of singularities. That is, they have 
different {\it chronotopologies}. 
These can be visualized using the
singular foliations of $S^2$ as $[0,1]\times S^1$,
with all points of the $S^1$ identified at the two 
endpoints of the $[0,1]$ interval. 

For spacelike $b$, 
if the time flow is ``up" for the two triangles
on the left in FIG.~\ref{fig:2D_discrete}, it is ``down" 
for the two triangles on the right. 
The $[0,1]$ factor is spacelike, two of these
spacelike slices being the two $a$ edges.
The $ S^1$ in this case is timelike, 
and the time period shrinks
to zero at the two ends of the interval.
This resembles the foliation for the 
smooth 2D sphere with two horizon points,shown in FIG.~\ref{fig:2D_continuum}.

For timelike $b$, the time flow is ``up'' for all four 
triangles in FIG.~\ref{fig:2D_discrete}.
Now it is the $[0,1]$ factor that 
is timelike, stretching from one $B$ vertex to
the other, while the
$S^1$ factor is spacelike.
One of the $S^1$ slices is composed of the two 
$a$ edges, joined at their endpoints.
The length of the spatial 
circles decreases when moving towards 
either of the $B$ vertices and reaches zero at these vertices.  
This triangulation is thus a 2D universe 
with spatial topology $S^1$ and with big bang and big crunch singularities at the $B$ vertices.

Due to the (Lorentz signature) Gauss-Bonnet theorem \cite{Chern63,Sorkin:2019llw} one can expect the Einstein-Hilbert actions of the spacetimes shown in FIG.~\ref{fig:2D_continuum} and FIG.~\ref{fig:2D_discrete} to be the same, as they have the same topology. So in any of the causal configurations above one should obtain the same action, namely $\pm\imath4\pi$ --- with the sign  a choice that we will discuss in \S\ref{ssec:Regge_intro}. Although the discretized spacetime is non-smooth one can still define its action via a discretized version of the Einstein-Hilbert action. The Regge action (\textit{cf.} \S\ref{ssec:Regge_intro}) is one such candidate and is consistent with the Gauss-Bonnet theorem, so the claims above can follow. 

\begin{figure}[h!]
    \centering
    \begin{subfigure}[c]{1\columnwidth}
        \centering
        \includegraphics[width=\linewidth]{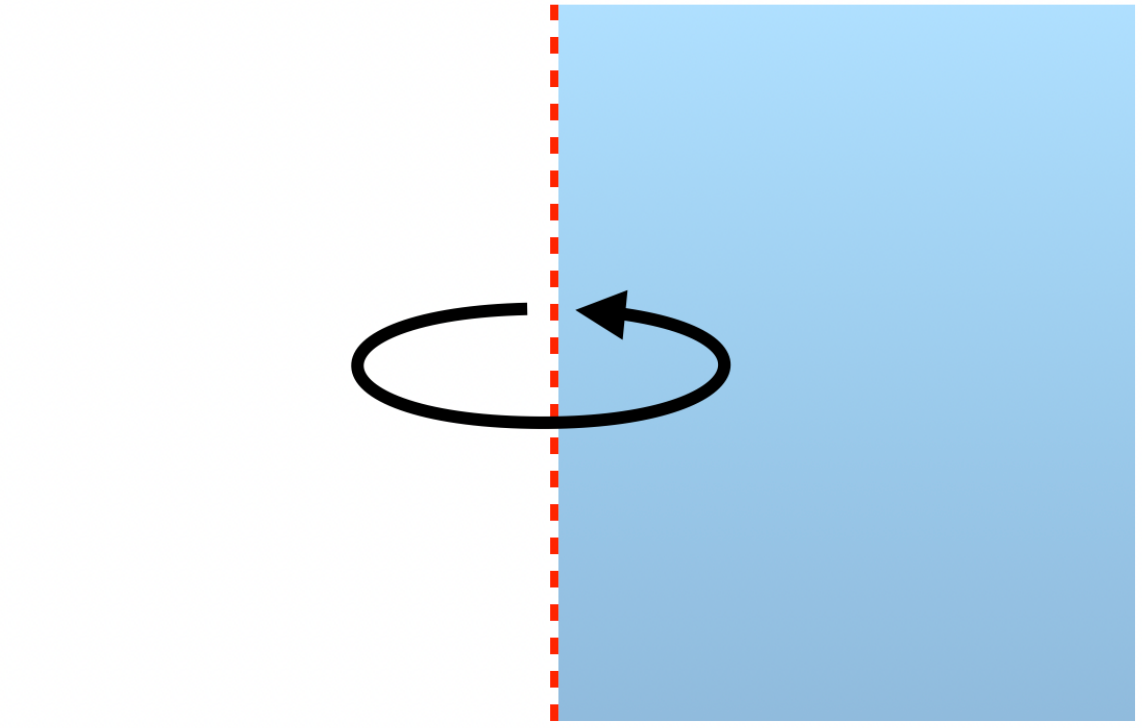}
        \caption{Foliation of a de-compactified spacetime 3-sphere (\textit{i.e.} $\mathbb R^3$) using a de-compactified spatial 2-ball (blue/right semi-plane) rotated around its boundary, a de-compactified 1-sphere (red/dashed line) which is a CTC singularity corresponding to the \textit{horizon}. }
        \label{fig:3D_continuum}
    \end{subfigure}
    \hfill
    \begin{subfigure}[c]{1\columnwidth}
        \centering
        \includegraphics[width=\linewidth]{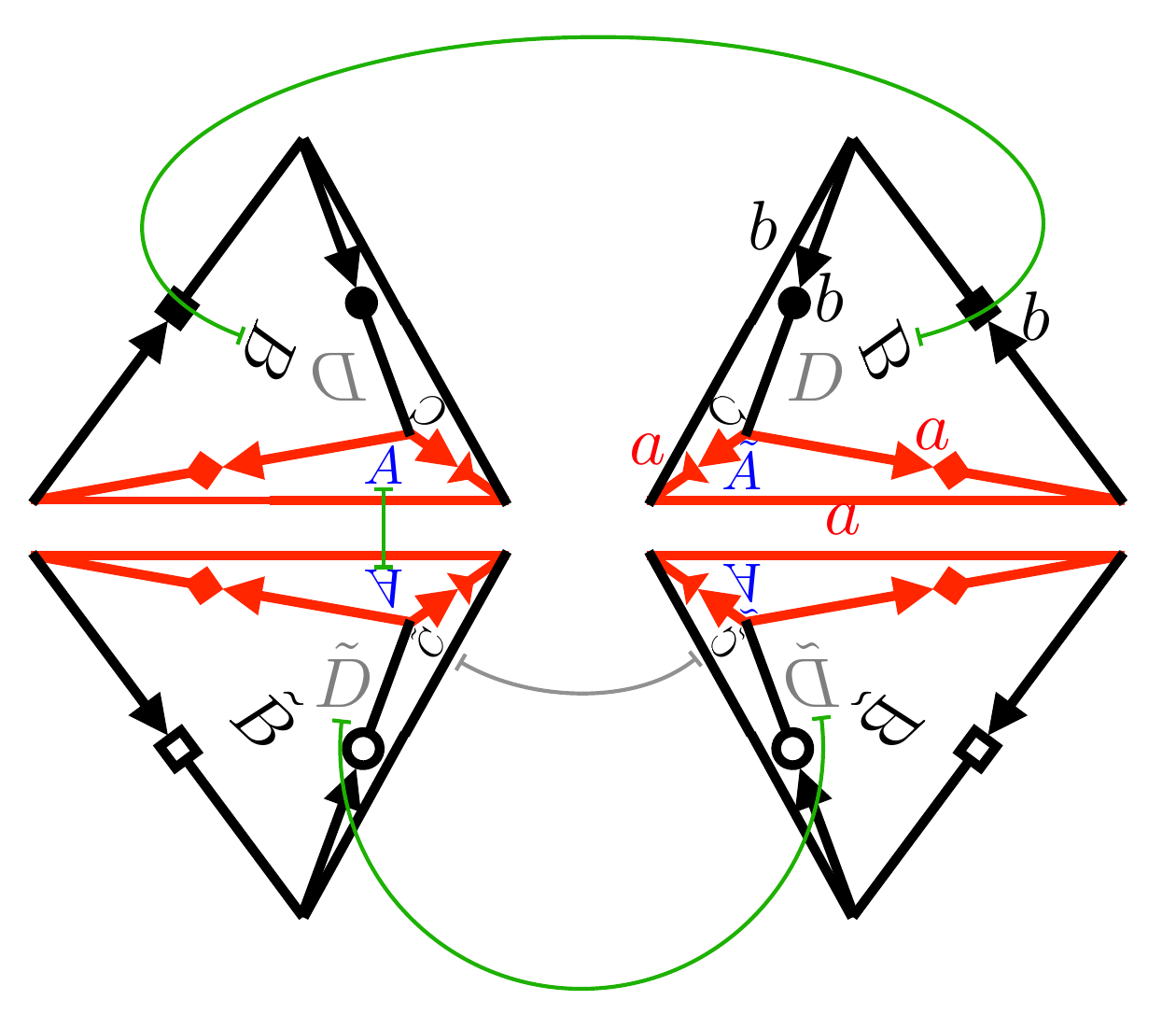}
        \caption{Triangulation of the spacetime depicted on the top figure, in which we glue four tetrahedra along their edges as instructed by the (oriented) arrows as well their faces/triangles as instructed by the upper-case letters (note that the base triangles $A$ and $\tilde A$ are only glued along their boundary edges) —see also the external lines. The horizon is replaced by the red perimeter (boundary of $A$ or $\tilde A$) and the metric by the lengths of the segments $a$ and $b$.}
        \label{fig:3D_discrete}   
    \end{subfigure}
    \caption{ Discrete and continuum 3-sphere spacetimes with   chronotopology of configurations in the gravitational partition function that computes the dimension of the Hilbert space of a 2-ball of space.
    }
    \label{fig:3D_spheres}
\end{figure}

To construct a  three-dimensional triangulation similar to the two-dimensional one we use four Lorentzian tetrahedra, see FIG.~\ref{fig:3D_discrete}. All tetrahedra have the same geometry determined by their six edge lengths squared
\be
(s_{12},s_{13},s_{14},s_{23},s_{24},s_{34})\,=\,(s_a,s_a,s_b,s_a,s_b,s_b).
\ee
The edges of type $a$ are spacelike.They bound two equilateral\footnote{Note that equilateral triangles are necessarily spacelike.} triangles, each of which represents ``space at one time'' in the partition function, and form the potential horizon.
The Lorentzian generalized triangle inequalities \cite{Tate:2011rm,Asante:2021zzh} furthermore demand that $3 s_b < s_a$. ({Here we exclude degenerate tetrahedra with vanishing volume.) Taking the triangle $(a,a,a)$ as base, we can introduce the height 
for the tetrahedra,
\begin{equation}
    s_h=s_b-\tfrac{1}{3}s_a.
\end{equation}
The Lorentzian generalized triangle inequalities  can then be expressed as $s_a>0$ and $s_h<0$. 

Let us discuss the light cone structure for this three-dimensional triangulation. We first note that a pair of (Minkowski-flat) $d$-dimensional simplices that are glued along a shared $(d-1)$-simplex can always be isometrically embedded into Minkowski space. In particular, for our triangulation  all points in the interior of the tetrahedra and in the interior of the triangles are light cone regular. Light cone irregularities can only appear at edges. To discuss these one considers the orthogonal projection of the piecewise flat geometry around the edge onto the two-dimensional piecewise flat geometry orthogonal to the edge.  This two-dimensional geometry will be Lorentzian if the edge is spacelike and Euclidean if the edge is timelike. Light cone irregularities at inner points of the  edge can thus appear only at a spacelike edge. Restricting to this case, note that the edge itself is projected to a vertex in this two-dimensional Lorentzian piecewise flat geometry.  We can therefore consider whether the vertex resulting from the projection is light cone regular with respect to this two-dimensional geometry. If this is not the case, points in the interior of the (spacelike) edge are not light cone regular with respect to the three-dimensional geometry and we will refer to such an edge as a light cone irregular edge.

As we will discuss in more  detail in \S\ref{ssec:Regge_intro}, these types of light cone irregularities will lead to imaginary contribution to the (Regge) action as well as branch cuts. Further types of light cone irregularities might appear at the vertices of the three-dimensional triangulation \cite{LollJordan,Asante:2021phx}, but will not contribute imaginary terms to the Regge action.

 For our three-dimensional triangulation we distinguish between three regimes. Namely, $(i)$ the $(abb)$ triangles (and therefore the $b$ edges) are spacelike, $(ii)$ the $(abb)$ triangles are timelike, but the $b$ edges are spacelike and $(iii)$ the $b$ edges (and therefore the $(abb)$ triangles) 
 are timelike. (The cases that either the $b$ edges or the $(abb)$ triangles are null do not contribute to the path integral.) The case $(i)$  appears for the range $0> s_h>-s_a/12$ and $(ii)$ for the range $-s_a/12 >s_h>-s_a/3$ whereas we have case $(iii)$ if $s_h<-s_a/3$.

Let us start with the case $(i)$ where the absolute value of the height $s_h$ is small enough so that the triangles $(a,b,b)$ are spacelike. The dihedral angle between the triangle $(a,b,b)$ and the triangle $(a,a,a)$ corresponds to a ``thin" Lorentzian angle between two spacelike planes, \emph{i.e.} an angle that does not include a  light cone. We are gluing four such ``thin" angles around the edge $a$, none of which contains a light cone. Thus, the edges of type $a$ are light cone irregular. In fact, these edges are CTC singularities. 

On the other hand, the $b$ edges are light cone regular: the dihedral angle at the $b$ edges is between two spacelike triangles and is ``thick", that is, it includes a light cone. (If the height is very small as compared to $s_a$,  we have an almost degenerate tetrahedron and the angle is almost half a full plane angle.) We glue two tetrahedra around the $b$ 
edges, and have thus two light cones at each inner point of these edges.

Increasing the absolute value of the height we hit the point where the triangles $(a,b,b)$ are null and then move into the regime $(ii)$ where these triangles are timelike.  
The dihedral angle between the two triangles $(a,b,b)$ and $(a,a,a)$ changes from a thin Lorentzian angle between two spacelike 
planes to a Lorentzian angle between a spacelike plane  
and a timelike plane. Thus it contains half a light cone. We glue four such angles around the edges of type $a$ —these $a$ edges are therefore light cone regular. In contrast to that, the dihedral angle at the spacelike $b$ edge in regime $(ii)$ corresponds to a ``thin" Lorentzian angle between two timelike planes, which contains only timelike directions. 
We glue two such angles together, and have thus only timelike directions in the plane orthogonal to the $b$ edge.  
These $b$ edges represent final or initial singularities, where timelike trajectories cannot be further extended.  

Increasing the absolute value of the height even further we reach the regime $(iii)$ where the $b$ edges become timelike. The $a$ edges are light cone regular, with the same reasoning as for case $(ii)$. The $b$ edges are now timelike and are therefore light cone regular.
The initial or final singularities, which in the case $(ii)$ were located along the $b$ edges, are now reduced to the two vertices where the $b$ edges meet.  We thus have a big bang to big crunch spacetime. 

\section{The gravitational (Regge) path integral}
\subsection{Introduction to (quantum) Regge calculus \label{ssec:Regge_intro}}}

With our triangulation defined, we proceed to study the gravitational path integral based on this triangulation. We need to start by adapting the Einstein-Hilbert action to it, which we do by means of the Regge action. We begin with a short review of the Euclidean and then Lorentzian Regge action that will help us understand the calculation in Section \S\ref{ssec:Regge_action}. A much more detailed discussion of the complex Regge action, which unifies the Euclidean and Lorentzian versions, can be found in \cite{Asante:2021phx}.

The Regge action \cite{Regge:1961px,Sorkin:1975ah} provides a \textit{discretization} of the Einstein-Hilbert action based on triangulations,\footnote{In principle one can loosen this restriction and work with more general polytope cellular decompositions.} usually considered to be made of piecewise flat simplices\footnote{One can also work with homogeneously curved simplices, whose geometry is also fixed by the edge lengths \cite{NewRegge}. This reduces discretization artifacts if one does have a non-vanishing cosmological constant \cite{Improved}. In three dimensions one  even obtains  triangulation invariant  and discretization artifact free results. But it has the disadvantage of leading to very involved expressions for the volume of the homogeneously curved simplices, and therefore the action.} whose geometry is uniquely fixed by the lengths of the edges. The variables discretizing the metric are therefore given by all of these lengths.\footnote{This is the case in length Regge calculus; other versions work with areas \cite{AreaRegge} or with areas and angles \cite{AreaAngle}.} Thus, the gravitational path integral is replaced by an integral over length assignments of the exponentiated `action' $e^W$. That is,\footnote{This expression is only pictorial. Note, for example, that in it we have not so far instructed how the sum over lapse signs would be implemented.}
\begin{equation}
    Z_\text{EH}=\int {\cal D}g \,e^W\longrightarrow\int \prod_e \mathrm d l_e \mu(\{l_e\})\,\, e^W.
    \label{eq:SOH_discretization}
\end{equation}
Here $\mu$ is a choice of measure that we will discuss below. For Euclidean quantum gravity one chooses $W=-S^\text{E}$, and for Lorentzian quantum gravity $W=\imath S^\text{L}$, with $S^\square$ the Regge action(s) in the corresponding signature.

There are several senses in which the Regge action is considered to \textit{discretize} the Einstein-Hilbert action. For example, in Euclidean signature, the solutions to the linearized discrete equations of motion set by the Regge action have been seen to converge in the continuum limit to the smooth Einstein linearized solutions when dealing with triangulations embeddable in a hyper-cubical lattice \cite{Rocek:1981ama,Rocek:1982tj,barrett1988convergence,Barrett:1988wd}. Likewise, on the non-perturbative side, it has been shown that Regge's curvature converges to Riemannian curvature \cite{Cheeger:1983vq,Feinberg:1984he}. These results are complemented by others in Lorentzian signature: For example in \cite{Sorkin:1975ah} it is shown that the Regge action is reproduced from the Einstein-Hilbert action when the manifold in question is taken to be an actual 
piecewise flat triangulation. (For a Euclidean version of this result see \cite{Friedberg:1984ma}.) It is also the case that several real time solutions to Regge's equations have been shown to approximate continuum solutions to Einstein's equations (\textit{e.g.} \cite{osti_4042676,Gentle_1998,Brewin:2000zh}).  Concerns that this behaviour might actually not be generic \cite{Miller_1995,Brewin_2000} have been addressed in \cite{Brewin:2000zh}. (See also \cite{Williams:1991pj,Misner1973} and references therein for a survey of Regge calculus.)

The key feature behind the Regge action is that one can define a notion of curvature localized on codimension-2 simplices (also known as \textit{bones}): the \textit{deficit angle} $\varepsilon$. This deficit angle\footnote{Although this is the standard name given to $\varepsilon$, in some situations ($\varepsilon<0$) it may not actually correspond to a \textit{deficit}, but to an \textit{excess}.} measures the failure for the part of the triangulation around its associated bone to be embedded into flat spacetime. This notion is most easily understood in Euclidean signature, so let us introduce it for this case first.

\begin{figure}
    \centering
    \includegraphics[width=0.5\textwidth]{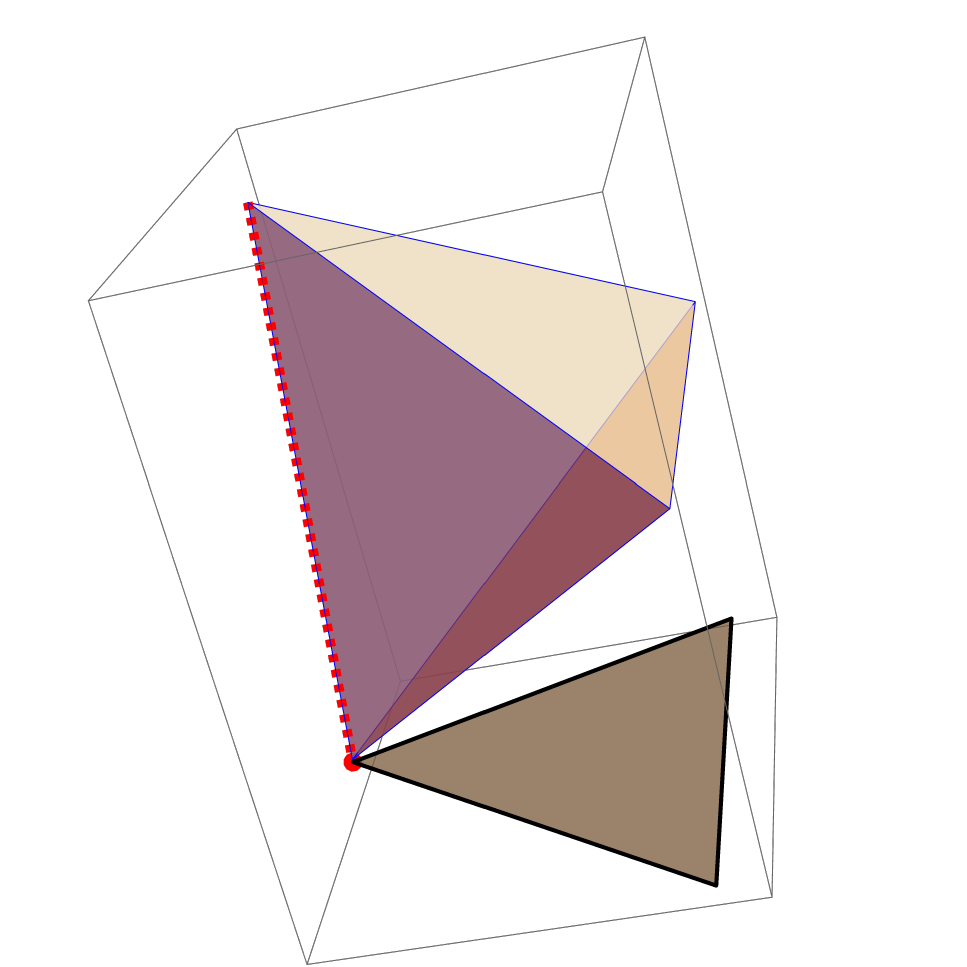}
    \caption{Three-dimensional example of the projection used to define the dihedral angle associated with an edge in a tetrahedron. By projecting out the dimension of the red/dashed edge, the tetrahedron becomes a triangle (in brown) and the edge a vertex. The (internal) 2D angle located at the vertex is the (internal) dihedral angle at its corresponding edge in the tetrahedron.}
    \label{fig:projection}
\end{figure}

To have an example in mind, suppose we consider three-dimensional space, as is done in \S\ref{ssec:Regge_action} (albeit in Lorentzian signature). Bones are then edges and attached to any of them there can be an arbitrary number of 3-simplices, that is, tetrahedra. Any edge in the bulk of the triangulation has a closed chain of tetrahedra $\tau$ glued around it and in each of them there is a dihedral angle located at the edge, which can be computed by projecting out the edge 
dimension so that each tetrahedron is mapped onto a triangle. Then the dihedral angle is the angle at the corresponding vertex in the resulting triangle (see FIG.~\ref{fig:projection}). If this chain of tetrahedra can be embedded into Euclidean space then the sum of these dihedral angles must give $2\pi$, otherwise there is a conical deficit angle $\varepsilon=2\pi-\sum_\tau \theta_{e,\tau}$.

\begin{figure}
    \centering
    \includegraphics[width=0.48\textwidth]{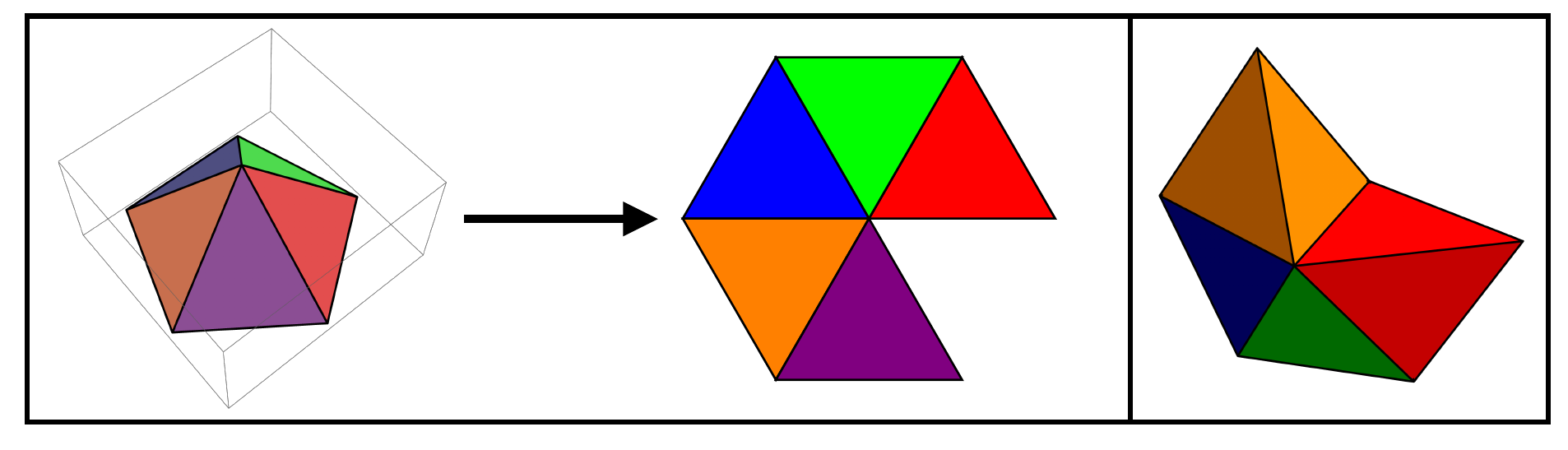}
    \caption{The deficit angle determines whether the simplices around a bone can be embedded in flat spacetime while remaining connected or not. In the left panel we have a positive deficit angle around a vertex, and therefore in order to embed the closed chain of triangles around, it must be broken. The right panel shows a broken chain for a three-dimensional triangulation.}
    \label{fig:deficit_angles}
\end{figure}

This picture can be generalized to any dimension, as bones are codimension-2 by definition, so the projection always results in a two-dimensional subspace. The two-dimensional case is particularly illustrative for understanding how the deficit angle encodes curvature, see for example FIG.~\ref{fig:deficit_angles}.

The Regge action hinges on this notion of curvature located at bones to capture the `curvature weighed by volume' essence of the Einstein-Hilbert action by discretizing the latter as follows:\footnote{One can also add a Gibbons-Hawking-York like term \cite{Hartle:1981cf}. However, as we deal with a triangulation without boundary, we omit its discussion.}$^,$\footnote{We work with units such that $8\pi G=\hbar=1$.}
\ba
 S^\text{E}_\text{EH}&=&-\frac12\int\mathrm d^D x\sqrt{g}(R-2\Lambda)\quad\quad
  \longrightarrow \nn\\
 S^{\text E}_\text R &=&-
 \hspace{-1em} \sum_{\beta\in\{\text{Bones}\}} 
 \hspace{-1em}     \text{Vol}_{D-2}(\beta)\,\varepsilon(\beta)+\Lambda 
 \hspace{-2em} \sum_{\sigma\in\{\text{$D$-Simplices}\}} 
 \hspace{-2em} \text{Vol}_D(\sigma)\,.\,\,\,
  \label{eq:Euclidean_action_discretization}
\ea
For a space of dimension $D=2$, if volumes of points are taken to be one,
the Regge curvature term 
in the action gives the same result 
as the curvature term in the continuum, which according to the
Gauss-Bonnet theorem depends only on the topology of the manifold.
While the definition 
\eqref{eq:Euclidean_action_discretization} is here motivated only
heuristically, it in fact has the convergence properties mentioned above.

The Lorentzian definition of the Regge action is very similar:
\ba
 S^\text{L}_\text{EH}&=&
 \frac12\int\mathrm d^D x\sqrt{-g}(R-2\Lambda)\quad\quad \longrightarrow \nn\\
 S^{\text L}_\text R &=&
 \hspace{-1em}
 \sum_{\beta\in\{\text{Bones}\}}
 \hspace{-1em}
 \text{Vol}_{D-2}(\beta)\varepsilon(\beta)-\Lambda
 \hspace{-2em}
 \sum_{\sigma\in\{\text{$D$-Simplices}\}}
 \hspace{-2em}
 \text{Vol}_D(\sigma)\,.\,\,
    \label{eq:Lorentzian_action_discretization}
\ea
(From here on we will drop the index $\text{R}$ from the Regge actions $S^\text{E}$ and $S^\text{L}$.)
Note that now bones may be null, timelike, or spacelike. All of their (dimension dependent) volumes are taken to be greater than or equal to zero.\footnote{An alternative possibility is to work with the square roots of the (signed) volume-squares. The signed 
volume-squares are negative for timelike building blocks. This alternative construction leads to the definition of a complex Regge action \cite{Asante:2021phx}. Adjusting for global factors of $\imath$, both definitions of Regge action are equivalent.}
In the Lorentzian case, the definition of dihedral angles needed for the deficit angles is however more involved, because when projecting out bone dimensions, the resulting 2-geometry may have a non-Euclidean signature. If the bone in question is timelike, then the resulting 2-geometry is Euclidean, hence the above definition applies; and null bones have zero volume, so do not contribute so the action.\footnote{
The role of configurations with null bones in the path integral is,
however, an open and interesting question.}
If the bone is spacelike, the 2-geometry resulting from the projection is piecewise Minkowskian flat. Thus, 
we need to understand how angles are defined 
in the Minkowski plane, which we now explain. 
The definition we shall give
is the one adopted in the works studying the continuum limit of the Lorentzian Regge action cited above. Further, it is such that angles are additive, and such that when the spacetime is two dimensional the Lorentzian Gauss-Bonnet theorem is satisfied, as stated in the previous section \cite{Sorkin:2019llw}.

\begin{figure}[ht!]
\begin{tikzpicture}[scale=0.6]
    \draw[gray!20] (-5,0)--(5,0) (0,-5)--(0,5);
    \draw[dashed, red,thick] (-4.5,-4.5)--(4.5,4.5) (-4.5,4.5)--(4.5,-4.5);

    \draw[thick,<-] (-4,-5) parabola bend (0,-3.5) (4,-5);
    \draw[thick,<-]  (4,5) parabola bend (0,3.5)(-4,5);
    \draw[thick,rotate=90 ,->] (-4,-5) parabola bend (0,-3.5) (4,-5) ;
    \draw[thick,rotate=90 ,->] (4,5) parabola bend (0,3.5)(-4,5);

    \draw [->] (0,0)--(3.8,2) ;
    \draw [->] (0,0)--(-3.8,2) ;

    \node[rotate = 45] at (5,3.2) {\scriptsize $+\infty $};
    \node[rotate = 45] at (3.5,5.3) {\scriptsize $+\infty \pm \imath \pi/2$};
    \node[rotate = -45] at (-3.5,5.3) {\scriptsize $-\infty \pm \imath \pi/2$};
    \node[below, rotate = -45] at (-5,3.9) {\scriptsize $-\infty \pm \imath \pi$};

    \node[below] at (4,0.5) {I};
    \node[left] at (0.5,4) {II};
    \node[below] at (-4,0.5) {III};
    \node[left] at (0.5,-4) {IV};

    \node[below] at (2.1,1.7) {$ a$};
    \node[below] at (-2.1,1.7) {$b$};

\end{tikzpicture}
\caption{Lorentzian angles provide boost parameters needed to take a ray into another. In order to boost a vector from one quadrant to another, one needs a complex parameter and correspondingly, each light cone crossing contributes with $\pm\imath\frac\pi 2$. In particular, the angle between the spacelike vectors $a$ and $b$, in quadrants I and II, respectively, has an imaginary part $\pm\imath\pi$, as in order to superpose them two light-rays need to be crossed.}
\label{fig:Lorentz_angles}
\end{figure}
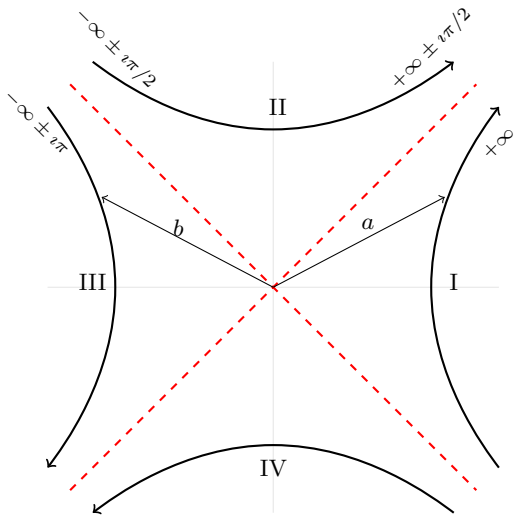

Just as a Euclidean angle is the one needed to rotate a unit vector into another in the plane, one can similarly define a Lorentzian angle as a boost parameter. A Lorentz boost (hyperbolic rotation) is implemented by the matrix
$$\Lambda_\eta:=\exp(\eta K)=\cosh\eta\,I + \sinh\eta\, K
$$
acting on the Minkowski components of vectors, where 
$K = \big(\begin{smallmatrix}  0 & 1\\ 1 & 0 \end{smallmatrix}\big)$.
However, there are no proper boosts taking spacelike vectors to timelike vectors and vice-versa, or relating space(time)-like vectors on sectors I and III (II and IV) of FIG.~\ref{fig:Lorentz_angles}.
For example, the boost taking the vector $\big(\begin{smallmatrix}  t \\ x \end{smallmatrix}\big)
=\big(\begin{smallmatrix}  0 \\ 1 \end{smallmatrix}\big)$
in sector I  of FIG.~\ref{fig:Lorentz_angles} to the light ray 
$\big(\begin{smallmatrix}  1 \\ 1 \end{smallmatrix}\big)$
has $\eta\rightarrow\infty$. On the other hand, $\eta$ {\it decreases} from $\infty$ to a
finite number in boosting from that light ray to, say, the vector $\big(\begin{smallmatrix}  1 \\ 0 \end{smallmatrix}\big)$, so the {\it net} boost angle is finite, and in fact equal to
zero in this case. 
For real $\eta$ the boost $\Lambda_\eta$ cannot map 
the vector $\big(\begin{smallmatrix}  0 \\ 1 \end{smallmatrix}\big)$ to the vector $\big(\begin{smallmatrix}  1 \\ 0 \end{smallmatrix}\big)$.
However,  with $\eta = \pm \imath\pi/2$, one has $\Lambda_{\pm \imath\pi/2} = \pm \imath K$,
which maps to $\pm\imath\big(\begin{smallmatrix}  1 \\ 0 \end{smallmatrix}\big)$,
\textit{i.e.} to the complex ray with the same ``direction" as
$\big(\begin{smallmatrix}  1 \\ 0 \end{smallmatrix}\big)$.
In this sense it is natural to extend to a definition of 
a generally complex boost angle between any two vectors in the Minkowski plane. 

In Euclidean signature the interior angle between two edges of a triangle is by definition positive, and to extend this to 
the Minkowski case we must adopt  sign conventions. 
We take the real part of the angle to be
the same as the sign of the Minkowski inner product of the two vectors
based at a vertex, and the imaginary part to be
$\pm \imath n\pi/2$, where $n=0,1,2$ is the minimal number of 
light-rays crossed to pivot one vector into another, and where
the sign before $\imath$ is a global ambiguity discussed below  \cite{Sorkin:1975ah,Sorkin:2019llw}. This definition of 
complex boost angle is consistent with additivity of angles \cite{Sorkin:2019llw}, and
with the analytic continuation of the Euclidean Regge action 
discussed below. 
As an example, consider the vectors $a$ and $b$ shown in FIG.~\ref{fig:Lorentz_angles}: Since $a\cdot b <0$, the real part of the angle is negative, and since 
two light-rays separate $a$ and $b$, the 
imaginary part is $\pm\imath\pi$.

The imaginary part of the angle has a sign ambiguity because any choice would lead to the same direction, \textit{i.e.} complex ray, of the boosted vector.
This can be linked with the ambiguity inherent in the definition $\imath:=\sqrt{-1}$ which in turn comes from the analytic structure of the square root. This will become more apparent in the framework of complex Regge calculus \cite{Jia:2021xeh,Asante:2021phx} discussed below. As we will discuss shortly, however,
when the light cone structure is regular the ambiguity is irrelevant and the Regge action is uniquely defined. 
The above cited work studying the continuum limit
for Lorentz signature considered only such 
light-cone-regular simplicial geometries.

Importantly, this definition implies that the Lorentzian angle covered by the whole Minkowski plane is $\pm\imath2\pi$, hence this imaginary angle replaces the $2\pi$ in the expression for the deficit angle $\varepsilon$ above. Therefore, if a bone's contribution to the Regge action is to be real, the dihedral angles attached to the bone must have a number of light-ray crossings ${\cal N}_c$ 
that exactly cancel $\pm\imath 2\pi$, \textit{i.e.} the light cone structure at the bone must be regular. If this is the case, the sign ambiguity of the angles' imaginary parts is not seen at the level of the action, since it `cancels'. Hence,  when dealing with light cone regular configurations, the Lorentzian action is real and uniquely defined.
If, however, ${\cal N}_c\neq 4$, then one has real parts in the path integral exponent of the form\footnote{This is in agreement with the imaginary contributions discussed in the continuum framework of \cite{Neiman:2013ap,Marolf:2020rpm,Colin-Ellerin:2020mva,Marolf:2022ybi}.
}
\ba\label{eq:imaginary_part}
    \text{Im}S^{\text L} \,=\, \pm 2\pi\text{Vol}_{D-2}(\beta)\left(1-\frac{{\cal N}_c}{4}\right),
\ea
where the area $\text{Vol}_{D-2}(\beta)$ is that of the irregular bone $\beta$. Thus, causally irregular histories are generically exponentially suppressed or enhanced, and the choice of sign  specifies which is the case for geometries with ${\cal N}_c<4$ and complementarily those with ${\cal N}_c>4$.\footnote{We remark that strictly speaking it is not the choice of sign for the angle's imaginary parts that determines which histories will be exponentially suppressed or enhanced, but rather their sign  relative to that of the `$\imath$' appearing in front of the action in the path integral's exponent.} Due to this ambiguity the Regge action has branch cuts along configurations with light cone irregular structure \cite{Asante:2021phx}. This can lead to an intricate topology of Riemann sheets, as discussed in detail in \cite{Asante:2021phx}. Nevertheless, if the light cone structure is regular, then the dihedral angles cancel the $\pm\imath 2\pi$, in which case the ambiguity associated with the Lorentzian angle branch cut is irrelevant, and the Regge action is analytic.

Instead of defining directly the Lorentzian Regge action as above, one can also obtain it by using a generalized Wick rotation. This is thoroughly derived in \cite{Asante:2021phx} (see also \cite{Sorkin:2019llw,Jia:2021xeh}). Here we will only sketch the procedure.

Starting from Euclidean space, one can introduce a generalized Wick rotation for the Euclidean time, $t_\text E\rightarrow e^{\imath\phi/2}t_\text L$, and show that the angle function has a branch cut when $\phi=\pm\pi$, corresponding to Lorentzian data. The resulting angle function agrees with the purely Lorentzian definition above up to a factor of $\pm\imath$.
Thus, choosing different sides of this branch cut leads to different signs for the light cone crossing contributions.

\begin{figure}
    \centering
    \includegraphics[width=0.5\textwidth]{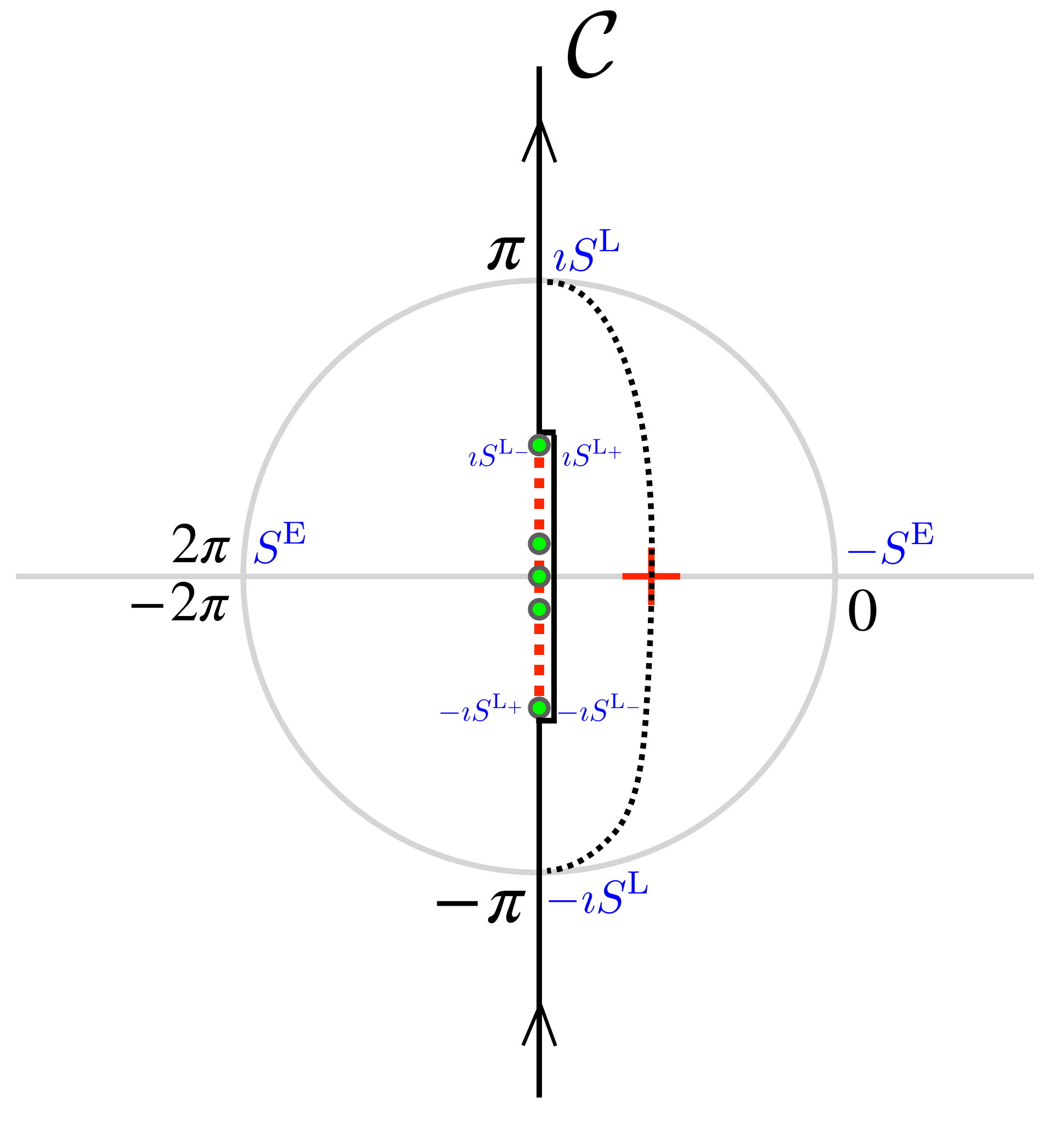}
    \caption{Illustration of some features of the analytically continued Regge path integral exponent $W$ 
    for a fixed value of $s_a$, as a function of the Riemann sheet variable $s_h$ (or equivalently the complex height $\sqrt{s_h}$). For specific values of the generalized Wick rotation angle $\phi$ it can reproduce the Euclidean and Lorentzian Regge exponents with either overall sign. For the triangulation under study there are branch points (green dots) at $s_h=0,\pm s_a/12,\pm s_a/3$ and the Riemann sheet we analyze has branch cuts (vertical red dashed lines) connecting them associated with light cone irregularities, where the Lorentzian action has imaginary parts. We have CTC singularities at the branch cuts from zero to $\pm s_a/12$. Depending on the side from which these are approached, one obtains Lorentzian Regge actions that  differ only in their imaginary part by a sign. We also show the original, and deformed contour of integration we use for our path integral for the case $s_a<8\pi\sqrt{3}/\Lambda$. The dotted portion of the deformed contour goes through a Euclidean saddle point (cross) and is a steepest ascent/descent flow-line of $\Re W$.}
    \label{fig:clock}
\end{figure}

This observation can be used to analytically continue the Regge action from the Euclidean regime to the Lorentzian one. More precisely, let us assume that a global generalized Wick rotation can be defined for the length-squared configuration space 
such that there is a Lorentzian portion of the complexified configuration space which is light cone regular. This applies to our triangulation: the height squared $s_h$ (\textit{cf.} \S\ref{sec:discretization}) can serve to determine a time variable and we have a regular regime where $s_b<0$ (\textit{cf.} \S\ref{sec:discretization}). Then one can analytically extend the Euclidean path integral exponent 
\ba
    W^\text E&=&-S^{\text{E}}
    \label{eq:Euclidean_ReggeW}
\ea
with $S^{\text{E}}$ defined in equation \eqref{eq:Euclidean_action_discretization}. The action as a function of the complexified $s_h$ is multi-valued due to the appearance of $\sqrt{s_h}$ —\textit{cf.} eq. \eqref{eq:W_a}. In order to integrate over both positive and negative height (to include both signs of the lapse) using the height squared $s_h$ as our variable, we need to extend the domain of $s_h$ to a double cover of the complex plane (Riemann `sheet'). This Riemann sheet can be parametrized by $r_h>0$ and $\phi\in(-2\pi,2\pi]$.\footnote{If there are causally irregular configurations along the Lorentzian lines at $\phi=\pm\pi$, then the extension above captures only a portion of the full Riemann surface. This is the reason why we refer to this partially extended domain as a ``sheet''.} We have two copies of the original $s_h$ plane and we recover the complex number $s_h$ \textit{via} the expression $r_h e^{\imath\phi}$.\footnote{Note that it is the domain of the height squared $s_h$ that is extended (in fact, doubled);  the domain of the height is not extended. In the context of the present paper one could have just worked in the height plane and avoided the Riemann surface extension. However, the latter seems to be the proper language for more general scenarios, because it allows one to deal more naturally with the branch cuts introduced by light cone irregularities. For example, steepest descent flows generically flow through these branch cuts, as illustrated in FIG.~\ref{fig:flows}, and therefore numerical techniques such as the gradient flow method will generically require one to consider several Riemann sheets. We also point out that working with length-squared variables is akin to working with metric (instead of \textit{vielbein}) variables.} Despite appearances, in the Riemann sheet context one is \textit{not} to identify $s_h$ values at $\phi$ with values at $\phi+2\pi$, as they correspond to heights ($\sqrt{s_h}$'s) of different sign. In this domain $W$ has the following behaviour 
(see also FIG.\ \ref{fig:clock}) \cite{Asante:2021phx}:
\begin{equation}
    W(\phi)
    \longrightarrow
    \begin{cases}
        +S^\text{E} \q\,\,\,\,\, \text{for} \,\,\,  {\phi\rightarrow -2\pi},\\
        -\imath S^{{\rm L}{}_+}\q \text{for}\,\,\,\,  {\phi\rightarrow -\pi-\delta},\\
        -\imath S^{{\rm L}{}_-}\q \text{for}\,\,\,\,  {\phi\rightarrow -\pi+\delta},\\
        -S^\text{E}\q\,\,\,\,\, \text{for} \,\,\,  {\phi\rightarrow 0},\\
        +\imath S^{{\rm L}{}_+}\q \text{for}\,\,\,\,  {\phi\rightarrow +\pi-\delta},\\
        +\imath S^{{\rm L}{}_-}\q \text{for}\,\,\,\,  {\phi\rightarrow +\pi+\delta},\\
        +S^\text{E}\q\,\,\,\,\, \text{for} \,\,\,  {\phi\rightarrow 2\pi}.
    \end{cases}
    \label{eq:W_sectors}   
\end{equation}

Here 
the $\pm$ indices for the \text{L}orentzian actions distinguish the sides of the branch cut in the case of  light cone irregular configurations, and $\delta>0$ is considered infinitesimal and thus gives prescriptions on how the limits are to be approached. The two Lorentzian actions are related by complex conjugation, $S^{{\rm L}{}_-}=(S^{{\rm L}{}_+})^*$. If a given Lorentzian configuration is causally regular, there is no branch cut and the two actions $S^{{\rm L}_-}$ and $S^{{\rm L}{}_+}$ agree.
Notice that the path integral exponent $W(\phi)$ agrees at $\phi=\pm2\pi$ so the Riemann sheet is glued along this line.

In the path integral below, we will navigate the branch cut side such that histories with ${\cal N}_c<4$ are exponentially \textit{enhanced} and therefore those with ${\cal N}_c>4$ are correspondingly suppressed. Such a choice yields an exponentially enhanced result for our Lorentzian path integral, which is needed to capture the expected horizon entropy.
One could also choose the opposite, suppressing, side of the branch cut.\footnote{Another possibility is to exclude light cone irregular configurations from the path integral. Of course, in our context, configurations with CTC singularities are required by the very nature of the partition function being computed, and our choice of triangulation has inadvertently introduced configurations with other light cone irregularities as well. In other contexts, studying the path integral under refinements might help to decide whether to include light cone irregularities, or how to navigate the branch points they give rise to. Such refinements likely lead to additional light cone irregular configurations, and the refined path integral would then depend on how these configurations are treated. One would like to obtain some sort of invariance under refinements, as such an invariance is related to a discrete notion of diffeomorphism symmetry \cite{Dittrich:2011ien,Review2022}.
} In fact, this choice of the suppressing side has been implemented in \cite{Asante:2021phx} in order to compute a no-boundary wave-function for a de Sitter cosmology from a Lorentzian simplicial path integral. In that case the results were very close\footnote{We emphasize that for this to be the case one needed to include the light cone irregular configurations in the path integral.} to the Lorentzian continuum mini-superspace path integral computations by Feldbrugge 
\textit{et al.} \cite{Feldbrugge:2017fcc}, which found an exponentially suppressed result for the wave function. Note that the opposite---that is, an exponentially enhanced result---was found by Diaz Dorronsoro \textit{et al.} \cite{DiazDorronsoro:2017hti}, also from the Lorentzian continuum mini-superspace path integral. This hints towards the fact that also in the continuum there exists a (hidden) choice for the Lorentzian path integral. We will comment more on this point further below.

Modulo the topological case when $D=2$ and $\Lambda=0$, the works studying the continuum limit of the real time Regge action have not discussed the case ${\cal N}_c\neq4$ and thus cannot be used to fix this ambiguity based on a classical criterion.\footnote{Let us remark that there is however a point of contact between our discussion and \cite{Sorkin:1975ah} with the continuum corner terms discussed in \cite{Neiman:2013ap}, but there the same ambiguity is faced.} However, a compelling criterion appears at the quantum level, 
for the integral over quantum field fluctuations
to converge. As spelled out long ago by Halliwell and Hartle \cite{HalliwellContours} in the context of the wave function of the universe, 
if quantum field theory of a scalar field in curved spacetime 
is to be recovered by expanding around a dominating saddle point of the path integral
with complex lapse function $N$ in the metric, it must be 
that $\Re N>0$. Otherwise, the fluctuation wave functional would not be not be normalizable, and one would
not recover the local vacuum of the quantum fields on the semiclassical spacetime background.\footnote{In \cite{HalliwellContours} this
criterion was stated as $\Re \sqrt{g} > 0$, which is the same
as $\Re N>0$ for a metric that is real except for the lapse,
but is otherwise weaker than the general condition required for 
a massless scalar field. Generalizing this criterion 
beyond scalar fields to $p$-form fields, 
Witten \cite{Witten:2021nzp} explored a selection criterion for complex metrics, following previous work in flat complex spacetimes \cite{Kontsevich:2021dmb}, and tested whether it rules out pathological examples and admits putative saddles that appear well motivated on physical grounds.} 
Moreover, even within pure general relativty with 
no additional fields, the convergence of the integral over graviton fluctuations alone imposes the same condition on the lapse \cite{Feldbrugge:2017mbc}. This amounts
to a consistency condition for the gravitational
effective field theory to provide a reasonable approximation
to an underlying, stable, UV complete theory.\footnote{We note, 
however, that this criterion is perturbative in nature,
and we cannot rule out the possibility that the full,
nonperturbative theory does not require this criterion.}
We shall refer to this complex metric selection principle
as the {\it fluctuation convergence criterion}, or just
{\it convergence criterion} where no confusion should arise.

In a setting somewhat closer to ours, in
the context of topology changing spacetimes in two dimensions, 
Louko and Sorkin
\cite{Louko:1995jw} proposed the same criterion as that in \cite{HalliwellContours}. The topology change 
necessitates the presence of points with light cone irregularity.
Deforming the metric infinitesimally into a smooth but complex one, they 
noted that there are two options for the deformation, leading to opposite signs for the imaginary part of the action, and 
argued that if one is to consistently couple a free, massless scalar field with these complex geometries, spacetimes with ${\cal N}_c>4$ must be suppressed in the sum over histories, as happens with our convention above, otherwise the variance of the Gaussian amplitude has a negative real part so the integral over fluctuations diverges.
Note that in this context the convergence criterion is applied not only at a saddle point, but at any configuration. While
the criterion has mostly been discussed in the context of its application at a saddle configuration---where it is clearly required if the saddle is to provide a consistent semiclassical approximation---it appears reasonable to apply it for all configurations.
We shall do so in the slightly weaker form of a ``non-divergence criterion'', \textit{i.e.}, we allow for the marginal Lorentzian case $\Re N = 0$, since the
complex Gaussian integrals that arise in that case are presumably tamed when computing physical observables.

\vspace{3mm}

\subsection{The Regge action \label{ssec:Regge_action}}

Having introduced the Regge calculus basics, we can finally move on to computing the Regge action of our particular triangulation.

We compute the Lorentzian Regge action of the triangulation in FIG.~\ref{fig:3D_discrete} as described in the previous subsection, namely: We  first compute the Euclidean version and then analytically continue using a generalized Wick rotation for the height. We will therefore encounter branch cuts in the Lorentzian sector at the causally irregular configurations described in \S\ref{sec:discretization}, and for the integration contour we will pick the side of the branch cuts on which the convergence condition holds and the configurations with ${\cal N}_c<4$ are exponentially enhanced.

Due to the symmetry reduction implemented in our triangulation we have only two types of edges, $a$ \& $b$, and correspondingly only two types of dihedral angles, $\theta_a$ \& $\theta_b$. Likewise, all of our four tetrahedra have the same volume. Therefore, the Euclidean exponent \eqref{eq:Euclidean_ReggeW} takes the form:
\ba
 W^\text E=-S^\text E &=&3\sqrt{s_a}(2\pi-4\theta_a)+6\sqrt{s_b}(2\pi-2\theta_b)
 \nn\\
 && -4\Lambda\text{Vol}_3(\text{tetrahedron}),
    \label{eq:Regge_action_1}
\ea
where the factors in front of the lengths count the number of bones of the same type and those in front of the dihedral angles count how many tetrahedra are glued along each type. 

The volume term can be computed with elementary geometry and the next step is to introduce the height variable by the replacement of $s_b$ with $s_h+\frac13s_a$, followed by the evaluation of the dihedral angles, which can be done as follows: We first embed our tetrahedron in Euclidean space, which because of the symmetry reduction can be done at once, then we project along each of the edges  in order to compute the relevant two-dimensional angle. Now, to compute the latter we use the following formula that simplifies the analytic continuation \cite{Asante:2021phx}, namely:
\begin{equation}
    \theta(\vec A,\vec B)=\imath\log\left(\frac{\vec A\cdot\vec B+\sqrt{(\vec A\cdot\vec B)^2-(\vec A\cdot\vec A)(\vec B\cdot\vec B)}}{\sqrt{\vec A\cdot\vec A}\sqrt{\vec B\cdot\vec B}}\right),
    \label{eq:2D_angle_log}
\end{equation}
where $\Box\cdot\Box$ denotes the Euclidean product (not necessarily two-dimensional) and we take principal branches for the logarithm and square roots, and complete their domains to the whole complex plane through setting $\sqrt{-1}=\imath$ and $\log(-1)=\imath\pi$. This formula follows from the identity $\text{arccos}z=-\imath\log(z+\sqrt{z^2-1})$, as well as $\vec A\cdot\vec B=\sqrt{\vec A\cdot\vec A}\sqrt{\vec B\cdot\vec B}\cos\theta(\vec A,\vec B)$. (For a more thorough discussion, see \cite{Jia:2021xeh,Sorkin:2019llw,Asante:2021phx}.)

What remains now is to extend to the Lorentzian regime, which is done by identifying 
the height squared $s_h$ with a time variable (or rather its square) that will be subject  to the generalized Wick rotation:
\begin{equation}
    s_h\rightarrow r_h e^{\imath\phi}.
    \label{eq:generalized_Wick}
\end{equation}
By doing so we are effectively complexifying the height squared and can perform analytic continuation over it. 

Note that we can  understand the height (square) as a lapse (square) parameter in the following sense: In a homogeneous isotropic continuum spacetime one may gauge fix the lapse function to be a constant, global parameter measuring the proper time normal to the spatial slices.  Similarly, we can understand $s_h$ to parametrize the extent of our simplicial universe in the timelike direction.
The Wick rotation in $s_h$ corresponds therefore to a Wick rotation of a lapse square parameter. 
The discrete action depends on $\sqrt{s_h}$, hence we will adopt a range of $(-2\pi,2\pi]$ for the Wick rotation angle $\phi$ in (\ref{eq:generalized_Wick}), so that all values of $\sqrt{s_h}$ in the complex plane
are included. From here on $s_h$ should be understood as living on a Riemann surface  —see the discussion below eq. \eqref{eq:Euclidean_ReggeW}. The sign ambiguity of $\sqrt{s_h}$ corresponds to that of the lapse, and we identify $\phi=\pi$ with positive lapse and $\phi=-\pi$ with negative lapse.

From \eqref{eq:generalized_Wick} it also follows that the region satisfying the fluctuation convergence (or rather ``non-divergence criterion'') is the one with $\phi\in[-\pi,\pi]$, because there (and only there), upon taking the square root, the corresponding arguments are in 
$[-\frac{\pi}2,\frac{\pi}2]$, so that $\Re\sqrt{s_h}\ge0$.

From the general discussion in \S\ref{sec:discretization} (see also \cite{Asante:2021phx}) we expect that the \textit{Lorentzian} action features branch points at $r_h=s_a/12$ and $r_h=s_a/3$, where the $(a,b,b)$ triangles and $b$ edges are null, respectively.  Additionally we have a branch point at $r_h=0$ resulting from the appearance of $\sqrt{s_h}$ (\textit{cf.} FIG.~\ref{fig:clock}.).
Further, with the branches chosen for the logarithm and square root in \eqref{eq:2D_angle_log}, the standard Euclidean action corresponds to $\phi=0$, and there are branch cuts at $\phi=0$ and $\phi=\pi$ for $r_h<s_a/3$. We thus perform the analytic continuation of the action from the $\phi\in(0,\pi)$ and $r_h>s_a/3$ region. This leads to
\begin{widetext}
\begin{align}
    W = {}&6\sqrt{s_a}\left[\pi+2 \imath \log \left(\frac{e^{-\frac{\imath \phi }{2}}\sqrt{s_a}+ \imath  \sqrt{12r_h}}{\sqrt{e^{-\imath \phi } s_a+12 r_h}}\right)\right]+\nonumber\\
   &+4e^{\frac{\imath \phi }{2}} \sqrt{9 r_h+3 e^{-\imath
   \phi } s_a}\left[\pi+\imath\log \left(\frac{-s_a+6 e^{\imath \phi } \left(r_h+\imath \sqrt{r_h \left(3 r_h+e^{-\imath \phi }
   s_a\right)}\right)}{s_a+12 e^{\imath \phi } r_h}\right)\right]
   -\Lambda  e^{\frac{\imath \phi }{2}} s_a \sqrt{r_h/3}~ .
    \label{eq:W_a}
\end{align}
\end{widetext}
This function is, by construction, $4\pi$-periodic and analytic for $\phi\in[-2\pi,2\pi)$ and $r_h>0$ with the exception of branch cuts along the $\phi=\pm\pi$ lines going from $r_h=0$ to $r_h=s_a/12$ and from $r_h=s_a/12$ to $s_a/3$. These branch cuts correspond to the (edge) light cone irregular regime, the former with CTC's around the $a$ edges, and the latter with closed spatial curves around the spacelike $b$ edges. We therefore have branch cuts covering the full interval $0<r_h<s_a/3$, so for simplicity we may abuse language and speak of having a single branch cut in this larger interval for the lines $\phi=\pm\pi$. 
  
\begin{figure}
    \centering
\includegraphics[width=0.48\textwidth]{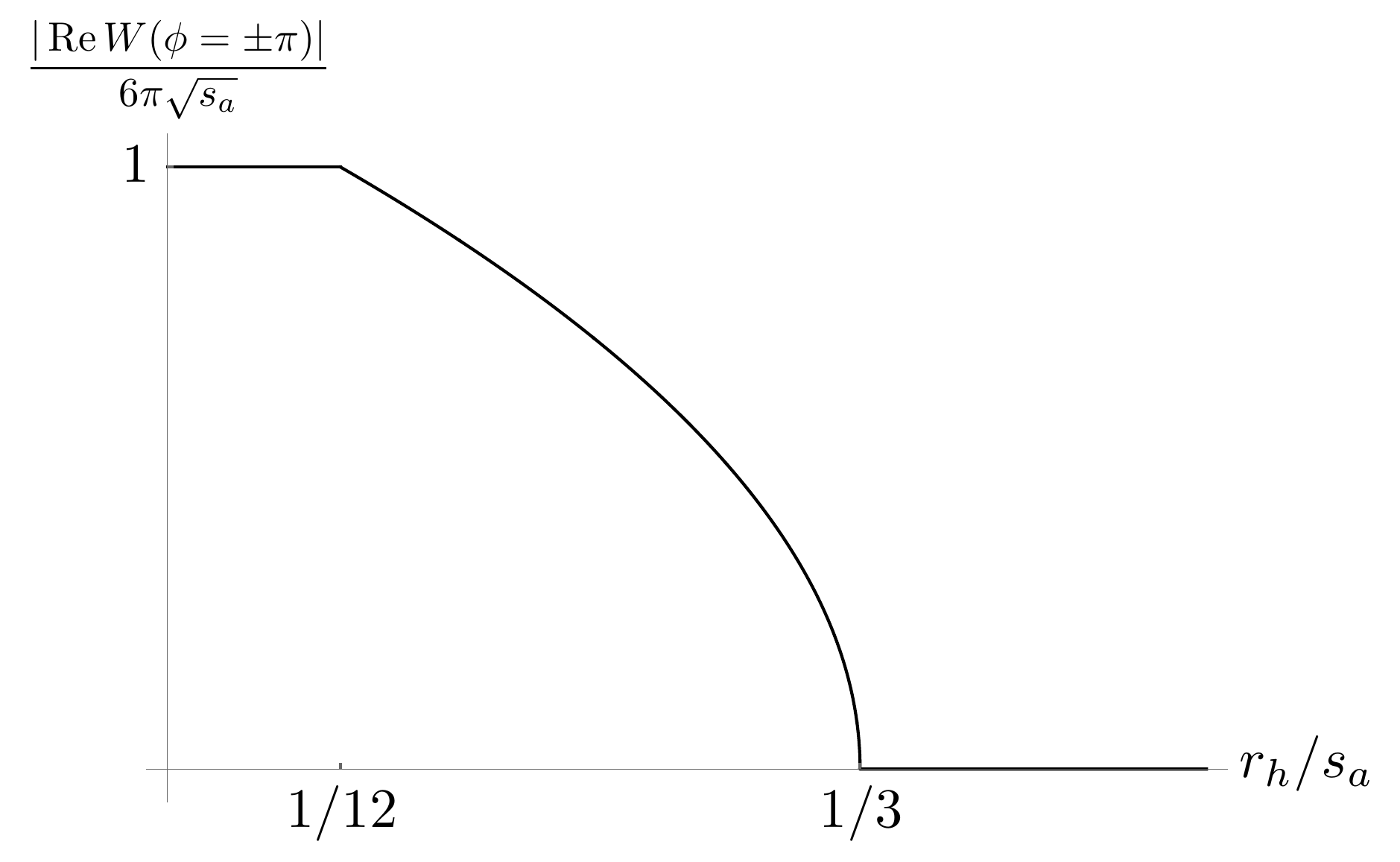}
    \caption{ The real part of $W$ along the Lorentzian lines is non-vanishing only along the branch cut, and is maximal and constant for the regime with CTC singularities.
    \label{fig:ReW_branch_cut}}
\end{figure}

Exponential enhancement of the partition function can arise only if the real part of $W$ is large. $\Re (W)$ vanishes at $\phi=\pm\pi$ for $r_h$ above the branch cut, and it develops a nonzero value next to the branch cut.
Crossing the branch cut the real part $\Re (W)$ changes sign. For $0<r_h<s_a/12$ we have $\Re (W)=\pm 6\pi \sqrt{s_a}$ at $\phi=-\pi \pm \delta$ and at $\phi=+\pi \mp \delta$. For $s_a/12<r_h<s_a/3$ 
we have $\Re (W)=\pm 12\pi \sqrt{s_b}$ at $\phi=-\pi \pm \delta$ and at $\phi=+\pi \mp \delta$. Note that $12\pi \sqrt{s_b}=12\pi\sqrt{e^{\imath\phi}r_h + s_a/3}$ 
as a function of $r_h$ at $\phi = \pm\pi$
is monotonically decreasing from $r_h=s_a/12$,
where it is equal to $6\pi \sqrt{s_a}$,  to $r_h=s_a/3$, where it is equal to $0$. The absolute value of the real part is thus maximal for the configurations with a CTC singularity, that is, for configurations with $r_h<s_a/12$.
A plot of this behavior is shown in FIG.~\ref{fig:ReW_branch_cut}.

As mentioned above, we will restrict to the branch cut side that agrees with the convergence criterion for the integration contour. Importantly, this
means that the partition function can receive exponential
enhancement from the light cone irregular regimes,
which is maximal from the configurations with CTC singularities.

More precisely, we take the path integral in question to be given by
\begin{equation}
    Z=\lim_{\epsilon\rightarrow0}\int_0^\infty\mathrm d s_a\mu_a(s_a)\int_{\mathcal C}\mathrm d s_h\mu_h(s_h,s_a) e^W,
    \label{eq:Z_path_integral}
\end{equation}
where we have $s_a\in(0,\infty)$ due to the triangle inequalities as discussed in \S\ref{sec:discretization}, and importantly the height-square integration is done over the black contour shown in FIG.~\ref{fig:flows}, for which we parameterized the height-square as $s_h=r_h e^{\imath\phi}$. Notice that the contour contains {\it both} Lorentzian branches $\phi=\pm\pi$, which correspond to positive and negative square roots of $s_h$ and, as explained above are associated with positive and negative lapse; this is needed in the continuum in order to impose the constraint and will play an important role in the next section. In order to navigate the branch cut in a way consistent with the convergence criterion, the contour is slightly deformed away from the Lorentzian lines for $r_h<s_a/3$ with the following $\epsilon$-prescription: The horizontal bottom line corresponds to an arc of radius $r_h=\epsilon$ going from $\phi=-\pi+\epsilon$ to $\phi=\pi-\epsilon$ that circumvents the $r_h=0$ branch point. Similarly, for $\epsilon<r_h<s_a/3+\epsilon$ the contour lies at $\phi=\pm\pi\mp\epsilon$. These branch cut portions are then joined to the Lorentzian lines $\phi=\pm\pi$ with arcs of radii $\epsilon$ going from $\phi=\pm\pi\mp\epsilon$ to $\phi=\pm\pi$, which we depict in FIG.~\ref{fig:flows} as the short horizontal lines. The contour continues along the Lorentzian lines from $r_h=s_a/3+\epsilon$ to infinity.

In eq. \eqref{eq:Z_path_integral} we have also split the quantum Regge calculus measure into two portions, one for what we will call the \textit{fixed-length} path integral $Z_{s_a}$, \textit{i.e.} that over $s_h$ at fixed $s_a$, and one for the remaining integral over $s_a$. 
The specific form of $\mu_a$ will not affect our discussion, as long as it does not overcome the large value of $e^W$ at the dominating semiclassical saddle. We also ask for $\mu_a$ to be positive and real, for reasons we discuss below.
With the motivation explained in the following paragraph we take the measure $\mu_h$ to be independent of $s_a$ and of the form
\begin{equation}
    \mu_h(s_h):=-\imath r_h^\alpha e^{\imath\alpha\phi}.
    \label{eq:measure}
\end{equation}
As shown in Appendix \ref{App:reality} the presence of the $-\imath$ factor and the reality of $\mu_a$ ensures that $Z$ is real, as it should be, because it computes the dimension of a Hilbert space.\footnote{In fact, using the saddle point approximation discussed below, or numerical evaluations, one can show that the overall sign on $\mu_h$ is such that $Z$ is positive.} As discussed below, we also require $\alpha<-\frac12$, in order for our integral to converge.

This form of the Regge measure can be motivated in different ways. For example, the case $\alpha=-3/4$ can be motivated as the discretization of the 
measure for the global lapse variable in a continuum minisuperspace path integral similar to
ours: The fixed $s_a$ integral  bears some resemblance with minisuperspace real time path integral computations of the no-boundary wave function \cite{Halliwell:1988wc,HalliwellContours,Feldbrugge:2017fcc,Feldbrugge:2017mbc,DiazDorronsoro:2017hti,Feldbrugge:2018gin,Marolf:1996gb}. 
In that setup one considers the lapse and scale factor as the only metric variables and computes the path integral corresponding to the transition amplitude for going from an `initial' scale factor $a_0$ to a `final' one $a_1$. The integral is over all intermediate scale factors and lapse. A variable transformation leads to an action quadratic in the scale factor variable, and thus a Gaussian path integral in this variable.\footnote{In the Gaussian integration the references above ignore the fact that $a^2$ is constrained to be positive, but see \cite{Halliwell:1988wc,Jia:2022nda} for discussions on this point.} Gauge fixing the lapse variable to a constant reduces the path integral further, to an integral over a global lapse only. Setting $a_0=0$ one obtains the no-boundary wave function depending on $a_1$. The global lapse integral is a close continuum analogue (albeit one spacetime dimension higher) of our fixed $s_a$ integrals. To see this note that a discrete version of the continuum computation would come about if one considers only the bottom tetrahedra (triangles) of FIG.~\ref{fig:3D_discrete} (FIG.~\ref{fig:2D_discrete}) and treats $s_a$ as a boundary variable analogous to $a_1$. Indeed, as mentioned before, a discrete version of this no-boundary calculation was performed in \cite{Asante:2021phx} and there the same Riemann surface topology as that of the current paper was found. In summary, our integral over $s_h$ can be compared with the continuum integral over global lapse. What is important for us is that the continuum integration of the scale factor produces a measure for the lapse integral, which as pointed out in \cite{Asante:2021phx} is discretized by a measure like \eqref{eq:measure} accompanied by a factor for $\mu_a$ that is positive and regular. Due to the fact that in order to obtain it in the continuum calculation one integrates over a scale factor, one could see this measure as capturing the information of some degrees of freedom that have been integrated out to go from an infinite dimensional minisuperspace integral over the scale factor, to a ``microsuperspace'' finite-dimensional integral over lapse.\footnote{
The references cited above derived the measure \eqref{eq:measure} for four-dimensional mini-superspace. A derivation for three-dimensional mini-superspace proceeds along the same line and leads only to a different numerical positive real factor, which we have absorbed into $\mu_a$.}

Another option for motivating this class of measures is to refer to three-dimensional quantum Regge calculus (without cosmological constant), for which one can derive a measure which leads to triangulation invariance\footnote{
A caveat here is that triangulation invariance might not hold anymore if one symmetry reduces the path integral, as we do in our case.
}
at one-loop order on a flat background, both for Euclidean and Lorentzian signature \cite{Meas1,Meas2,Meas3}. Apart from a phase and numerical pre-factors, the measure is given by $\mu=\prod_\tau V_\tau^{-1/2}$, where $V_\tau$ denotes the volume of a tetrahedron $\tau$. Our triangulation has four tetrahedra with height square $s_h$, the measure would therefore be $\mu\propto s_h^{-1}$, \textit{i.e.} $\alpha=-1$. We note, however, that $\mu_a$ is \textit{not} regular at $s_a=0$, and therefore, if used without modifications it would overcome the large value of $e^W$ in the semiclassical saddle when $s_a$ is small enough. In other words, this measure does not satisfy the conditions we put on $\mu_a$. This is in contrast to the divergence of $\mu_h$ for small $r_h$, which as we will see is not problematic for the validity of the saddle point approximation.

The way we have written $\mu_h$ in \eqref{eq:measure} is such that it has been analytically continued exactly as we did with $W$. We are therefore in effect introducing a Riemann surface coordinatized by $\phi$ and $r_h>0$ for the whole integrand in order to avoid the introduction of a possible additional branch cut. We remark that the measure in general is not $4\pi$-periodic. So, although $W$ is $4\pi$-periodic, our fixed-length integrand may not be. The integrand will however be periodic as long as $\alpha$ is rational.

Equation \eqref{eq:Z_path_integral} exemplifies the virtues of Regge calculus: namely, it possesses Einstein-like dynamics through the Regge action and gives a tractable finite dimensional integral, which as such can shed some light on quantum gravitational dynamics. Likewise it can be used as a lattice model for numerical evaluations. In particular, we can use it to explore some of the questions raised in \S\ref{sec:intro} and \S\ref{sec:CF}, as we are about to see.

\subsection{Saddle point approximation of the path integral \label{ssec:saddle_analysis}}

One such question was whether the (continuum) Lorentzian integration contour can be deformed in a way such that the Euclidean de Sitter saddle dominates. The first thing to check is therefore whether the discrete model has a Euclidean-dS-like saddle. 

Let us begin exploring this question by analyzing fixed-length saddle points, that is, points that for fixed $s_a$ extremize the exponent as a function of $s_h$. To see whether there are saddles along the Euclidean line at $\phi=0$ or at the Lorentzian line at $\phi=\pi$ we investigate the behaviour of $W$ for $r_h$ small or large compared to $s_a$. 
For small $r_h$ 
at $\phi=0$ and small positive $(r_h-s_a/3)$ at $\phi=\pi$
we have
\begin{widetext}
\begin{subequations}
\begin{eqnarray}
    W(\phi=0)& \sim & 6\pi \sqrt{s_a}- {s_a \Lambda}\sqrt{r_h/3}+ {\cal O}(r_h^{3/2}/s_a) \, , \label{eq:W_small_asymptotics_Euclidean}\\
    W(\phi=\pi)& \sim &\imath\left(-{\sqrt{s_a} (s_a \Lambda+18\log(3))}/{3} +  12 \pi \sqrt{r_h-s_a/3}\right) + {\cal O}((r_h-s_a/3)^{3/2}/s_a)\, .
\end{eqnarray}
\label{eq:W_small_asymptotics}
\end{subequations}
\end{widetext}
That is, (minus) the Euclidean action $W(\phi=0)$ has a decreasing behaviour for small (and growing) $r_h$, whereas the Lorentzian action $-\imath W(\phi=\pi)$ is increasing for small (and growing) $(r_h-s_a/3)$. Importantly this behaviour does not depend on the value of $s_a>0$. 

The asymptotic behaviour for large $r_h$ is given by
\begin{equation}
    W\sim  \frac{1}{\sqrt{3}} e^{\imath \phi/2}\Lambda\left(s_a^\Lambda-s_a  \right) \sqrt{r_h} +{\cal O}(r_h^{-1/2}s_a) \q ,
    \label{eq:W_asymptotics}
\end{equation}
and changes at the threshold value
\begin{equation}
    s_a^\Lambda:=8\sqrt{3}\pi/\Lambda\approx{43.5}/\Lambda.
    \label{eq:threshold}
\end{equation}
Thus we have for $s_a<s_a^\Lambda$ that $W(\phi=0)$ is decreasing for small $r_h$ and increasing for large $r_h$. The function has therefore at least one minimum. In contrast, we have that $-\imath W(\phi=\pi)$ is increasing for both small and large $(r_h-s_a/3)$, so there might
be no extrema. For $s_a>s_a^\Lambda$ the situation is opposite: there must be at least one maximum for the Lorentzian action, while 
there might be no extrema for the Euclidean action.
Indeed, numerical investigations show that there is exactly one saddle point along the $\phi=0$ line (and no Lorentzian saddles) for $s_a<s_a^\Lambda$ and exactly one saddle point along the $\phi=\pi$ line (and no Euclidean saddles) for $s_a>s_a^\Lambda$. 

\begin{figure*}
    \centering
    \begin{subfigure}[t]{0.47\textwidth}
        \centering
        \includegraphics[width=\linewidth]{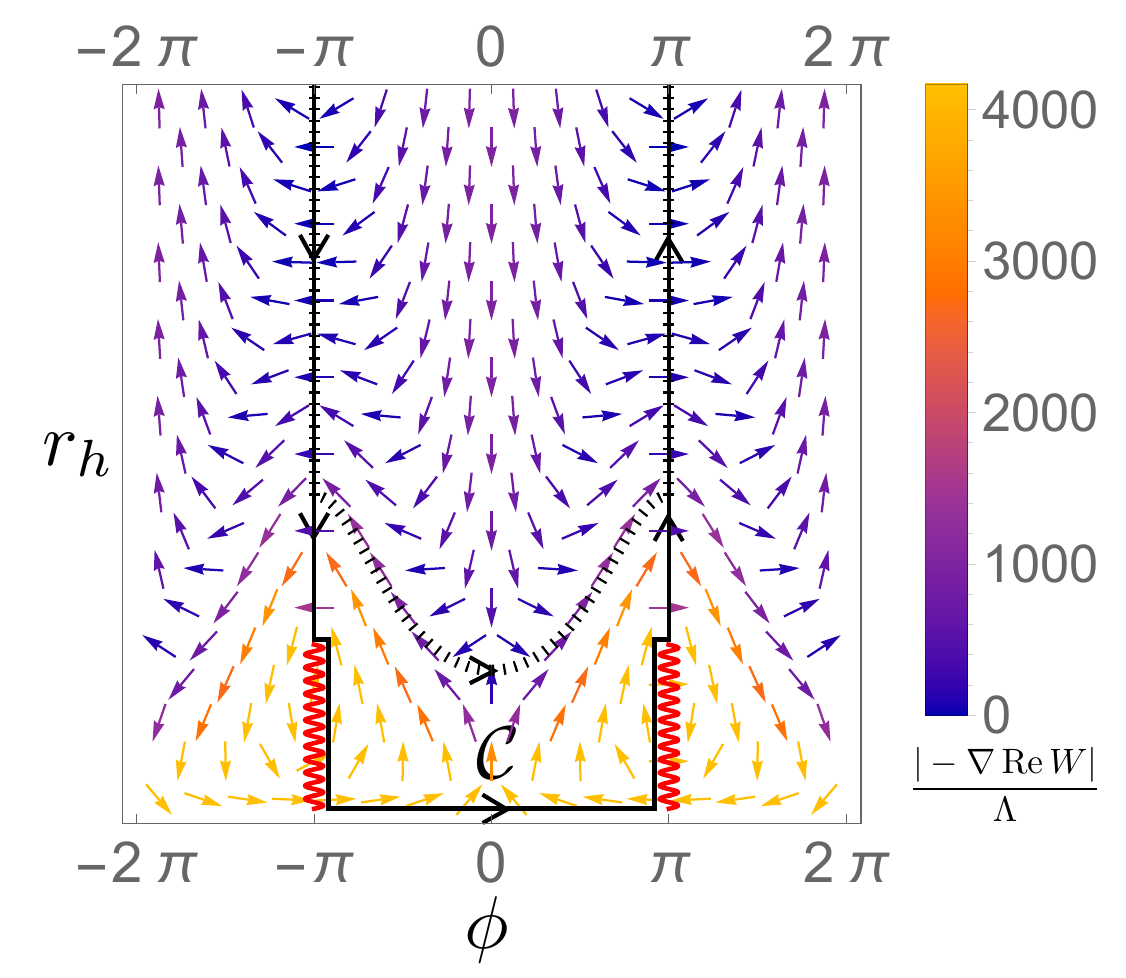}
        \caption{Steepest descent  flow of $\Re  W$ for the case of fixed $s_a=s_a^*<s_a^\Lambda$. In this case we have exactly two Euclidean saddle points of opposite sign for the lapse, and the contour can be deformed to the dashed one, which in the portion with $\phi\in(-\pi,\pi)$ agrees with the Lefschetz thimble of the $\phi=0$ Euclidean saddle.
         }
        
        \label{fig:Euclidean_saddles}
    \end{subfigure}
    \hfill
    \begin{subfigure}[t]{0.45\textwidth}
        \centering
        \includegraphics[width=\linewidth]{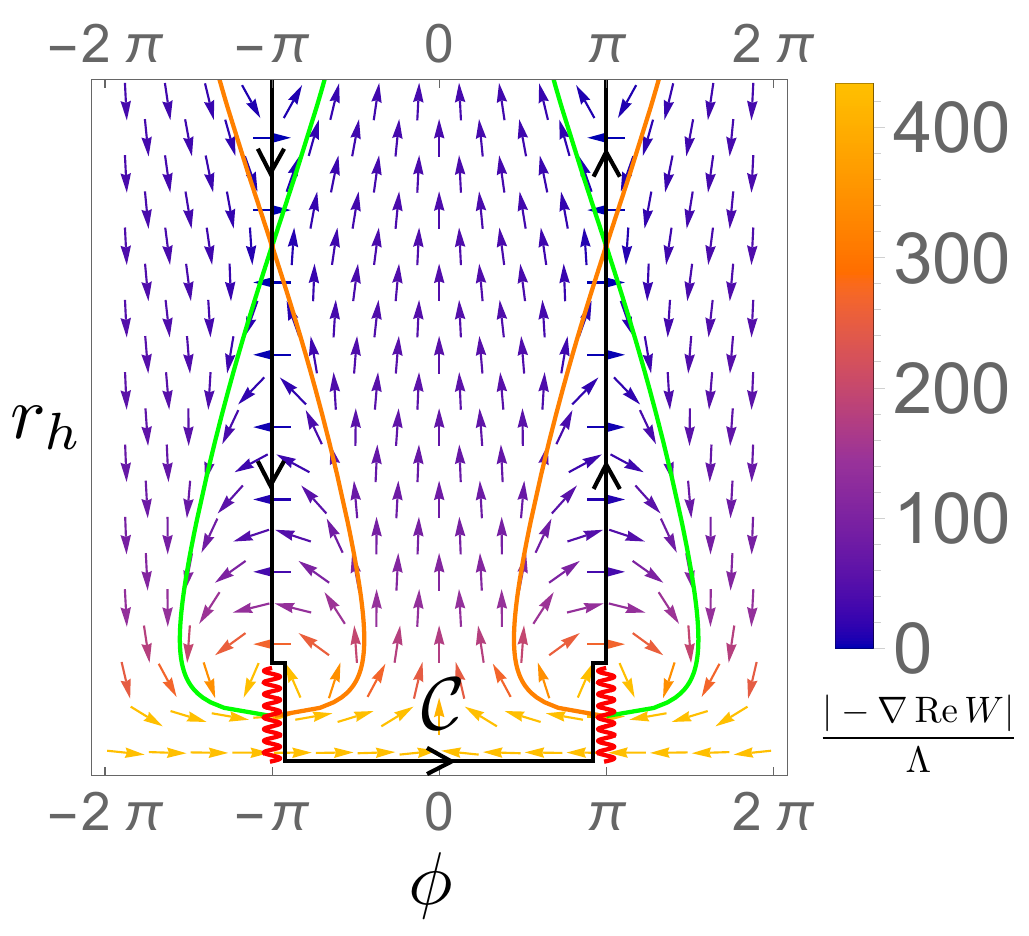}
        \caption{Steepest descent flow of $\Re  W$ for fixed $s_a>s_a^\Lambda$. Here we have exactly two Lorentzian saddle points of opposite sign for the lapse, and the contour can be closed at infinity using an (infinite) arc with $\phi\in(-\pi,\pi)$. We also show Lefschetz thimbles and anti-thimbles of the critical points. Note that each anti-thimble
        (orange, steepest ascent flow line) intersects the quasi-Lorentzian contour in two points, once at the saddle and once at the branch cut, which would have been impossible had $W$ not developed a positive real part at the branch cut.}
        \label{fig:Lorentzian_saddles}   
    \end{subfigure}
    \caption{Riemann sheet of the path integral exponent $W$ as a function of the variable $s_h$ (or equivalently the complex height $\sqrt{s_h}$) represented with coordinates $(\phi,r_h)$. Note that the lines at $\phi=-2\pi$ and $\phi=2\pi$ are to be identified and that the one at $r_h=0$ must be understood as a point (zero radius). The original ($\epsilon$-deformed Lorentzian) integration contour is shown as the solid and connected line, which circumvents infinitesimally the branch cuts (wavy red lines at the bottom) according to the convergence criterion. The contour includes both signs of the lapse, \textit{i.e.} both the lines $\phi=\pi$ and $\phi=-\pi$, as needs to happen in the continuum to implement the Hamiltonian constraint. Because the $r_h=0$  line on the diagram corresponds to a branch point, the horizontal portion of the contour is understood as an infinitesimal arc around the height-squared origin.}
    \label{fig:flows}
\end{figure*}

From equation (\ref{eq:W_sectors}) we see that $W(\phi=0)=-W(\phi=2\pi)$, and  $ \Im (W(\phi=\pi))=-\Im (W(\phi=-\pi))$. The  Euclidean and Lorentzian saddles appear therefore in pairs; \emph{i.e.}, if there is a saddle at $(r_h,\phi)$ then there is another saddle at $(r_h,\phi+2\pi\, \text{mod}\, 4\pi)$. A similar appearance of pairs is found in the mini-superspace discussion (see \textit{e.g.} \cite{DiazDorronsoro:2017hti}), where these pairs represent positive and negative lapse solutions. The two regimes, $s_a<s_a^\Lambda$ and $s_a>s_a^\Lambda$,  are shown in FIG.~\ref{fig:flows}.
The critical points of the flow shown in this figure, \textit{i.e.} points where $\grad \text{Re}\,W=0$, coincide  with the saddle points of $W$ (with $s_a$ kept fixed), thanks to the Cauchy-Riemann equations.
Note that, apart from the saddles discussed, there are no further saddles if we stay on the Riemann sheet shown in FIG.~\ref{fig:flows}, that is, if we do not cross any of the branch cuts.
%

The threshold behaviour mimics the continuum\footnote{It has also appeared in other closely related discrete analyses \cite{Dittrich:2021gww,Asante:2021phx}.} \cite{Jacobson:2022jir,Feldbrugge:2017kzv} and is related to the fact that topological hemispheres can solve the vacuum Euclidean Einstein field equations with cosmological constant 
as long as their boundary radius is smaller than the dS radius, and that there are complex saddles when it is larger. Thus, it could be that the Lorentzian saddles we find are discrete avatars of these continuum complex geometries. This also suggests that the threshold scale $s_a^\Lambda$ acts as the de Sitter radius squared.

In summary,
the partition function path integral can be split into two segments:
$s_a\in(0,s_a^\Lambda)$, for which each $s_a$ admits Euclidean saddles for the 
 $s_h$ integral; and $s_a\in(s_a^\Lambda,\infty)$, for which each $s_a$ admits Lorentzian saddles.

We are going to evaluate the integrals using deformations to different contours for the two different cases. But before discussing these contour deformations, we note that one can establish the convergence of the integral along the original contour in the limit of $r_h\rightarrow \infty$, as long as the parameter $\alpha$ in the measure satisfies $\alpha<-1/2$. To this end one uses Dirichlet's test for convergence of improper integrals, the details of which can be found in Appendix \ref{App:Dirichlet}.

We will now analyze the integral over $s_h$, for $s_a<s_a^\Lambda$, where we have Euclidean saddles.

In FIG.~\ref{fig:Euclidean_saddles} we show the steepest descent flow of the exponent's real part, $\Re  W$, for the case with $s_a\in(0,s_a^\Lambda)$. The figure indicates that the original contour can be deformed to the dashed one, which passes through the steepest descent contour of the $\phi=0$ saddle point (that is, the \textit{Lefschetz thimble}) to then rejoin the original Lorentzian contour. Along the thimble $W$ has positive real part: $\Re  W$ vanishes at $\pm\pi$ away from the branch cut, because there the simplicial geometry is (edge) light cone regular (\textit{cf. \S\ref{ssec:Regge_intro}}), so in the thimble, $\Re W$ ascends from zero to a positive maximum and then descends back to zero. In the rest of the deformed contour, $\Re  W$ is zero, because the contour goes along the Lorentzian lines. The thimble's contribution therefore dominates in a semiclassical expansion, where it can also be approximated by a saddle point evaluation. This conclusion is only strengthened when taking into account the measure as it suppresses larger heights.

In fact, by arguments similar to the ones we will give for the case with Lorentzian saddle points, the arcs at infinity from $\phi=-2\pi$ to $\phi=-\pi$ and from $\phi=\pi$ to $\phi=2\pi$, have vanishing contour integrals, so one could actually further deform the $r_a>s_a/3$ sub-portions of the contour along the flow all the way to the line at $\phi=-2\pi$ and the line at $\phi=2\pi$, which we \textit{should not} identify when taking into account the measure $\mu_h\sim s_h^\alpha$ for general $\alpha$. In this way, the left (right) sub-portion would become the portion of the Lefschetz thimble with $-2\pi<\phi<-\pi$ ($\pi<\phi<2\pi$) followed by the sub-portion of the Euclidean branch at $\phi=-2\pi$  ($\phi=2\pi$) that goes from its critical points up to infinity. 
Thus, the full integral is equivalently expressed as that over the full Lefschetz thimble---not just the portion with $\phi\in(-\pi,\pi)$---and the sub-leading Euclidean portions.
The Euclidean sub-portions are also steepest descent flow lines, and therefore their contributions are sub-leading  with respect to the full thimble associated with the $\phi=0$ saddle point, which goes from $\phi=-2\pi$ to $\phi=2\pi$.  
The full integral is manifestly convergent and, importantly, saddle point dominated.

In conclusion, for each value of $s_a\in(0,s_a^\Lambda)$, the fixed-length integral can be approximated, in a semiclassical expansion, by
\begin{equation}
    Z_{s_a}\sim \exp\bigl({W_\text{HJ}(s_a)}\bigr),
    \label{eq:fixed_length_HJ}   
\end{equation}
with $W_\text{HJ}$ the \textit{fixed-length Hamilton-Jacobi function, \textit{i.e.},} the exponent $W$ evaluated on the fixed-length saddle, $W(s_h(s_a),s_a)$.

\begin{figure}
    \centering
    \includegraphics[width=0.45\textwidth]{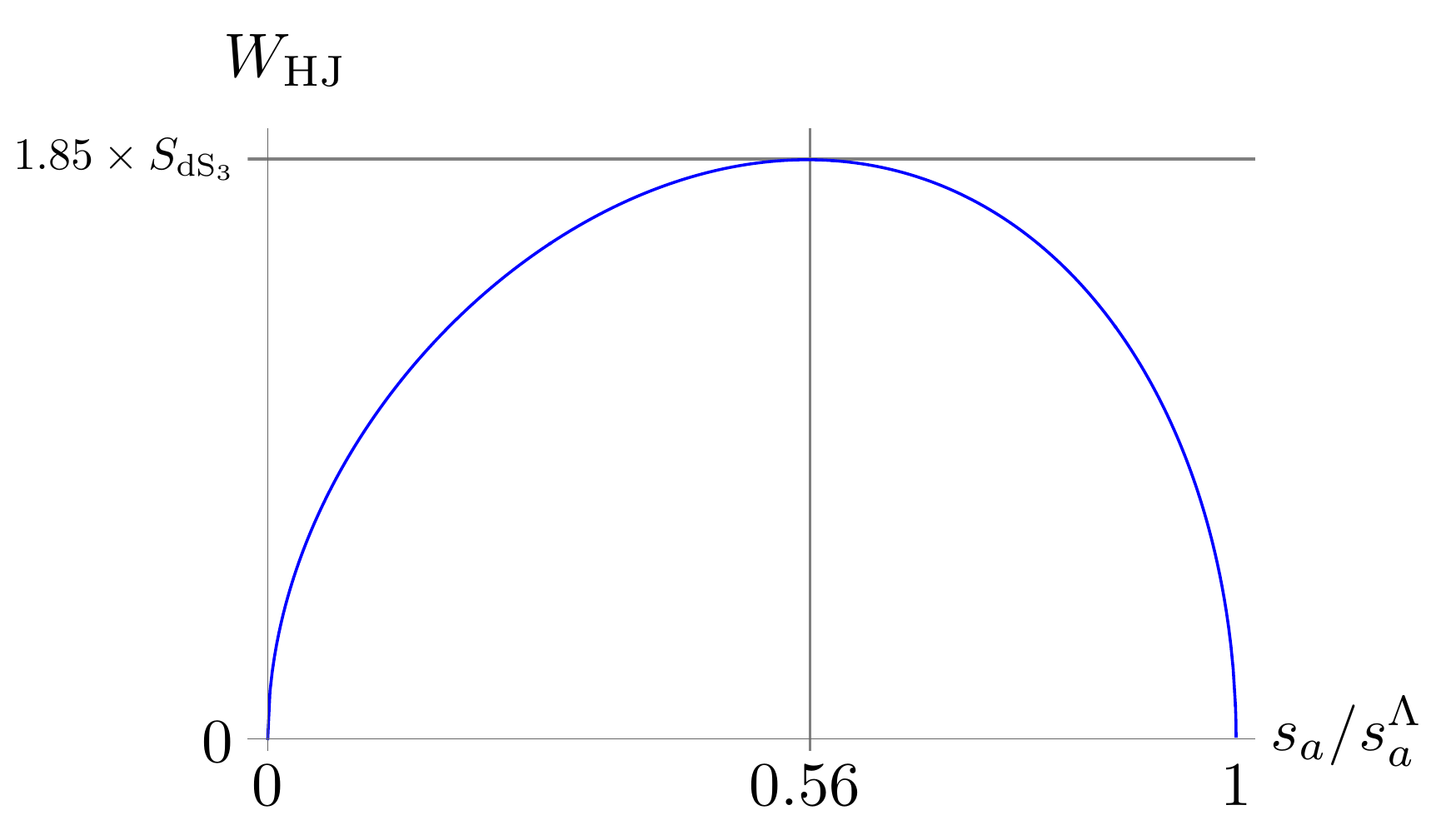}
    \caption{Fixed-length Hamilton-Jacobi function, $W_\text{HJ}(s_a)$ with $s_a<s_a^\Lambda$, that is for the range where only Euclidean saddle points exist. 
    \label{fig:fixed_length_HJ}}
\end{figure}

In FIG.~\ref{fig:fixed_length_HJ} we show the behaviour of $W_\text{HJ}$. It vanishes at $s_a=0$, increases up to a maximum at $s_a=s_a^*\approx24.3/\Lambda$, and returns to zero as $s_a$ approaches $s_a^\Lambda$. In this limit, $r_h(s_a)$ tends to $\infty$. On the other hand $r_h(s_a)$ goes to zero as $s_a\rightarrow0$, which can be argued by minimizing \eqref{eq:W_small_asymptotics_Euclidean} including the next-to-next-to-leading order term and noting that the resulting saddle is consistent with the small $r_h$ approximation. More specifically, we have saddles at $r_h\approx s_a^2\Lambda/216$ for $s_a\ll s_a^\Lambda$.

As hinted at in the previous section, one might worry that the divergent behaviour of $\mu_h$ as $r_h\rightarrow0$ changes the validity of using fixed-length saddle points of $W$ for semiclassical evaluations and therefore of \eqref{eq:fixed_length_HJ}, especially because 
at the saddle $r_h(s_a)\overset{s_a\rightarrow0}{\longrightarrow}0$. This is not the case. Although it is true that the contour integral receives 
locally large contributions from the region  near $r_h=0$, they evidently cancel each other, since the contour can be deformed away from this region. Moreover, the small $r_h$ divergence does not invalidate our saddle point approximation. To see that we note that when considering the steepest descent flow of $\Re(W+\log\mu_h)$ one still sees the same qualitative behaviour as in FIG.~\ref{fig:flows} in the $-\pi<\phi<\pi$ region, which indicates that the saddle point approximation is valid when using the saddle points of the \textit{joint exponent} $W+\log\mu_h$ (properly analytically continued). The position of these fixed-length saddles turns out to be always finite, approaching $r_h=(2\alpha\hbar G)^2$ as $s_a\rightarrow 0$, and therefore produces a finite exponent. If  $\hbar$ is sufficiently small compared to the action, the saddles of the joint exponent will still give the behaviour of FIG.~\ref{fig:fixed_length_HJ}, since the contribution of $W$ in the joint exponent will become dominant, as the action is divided by $\hbar$ whereas the measure adds a logarithmic dependence on $\hbar$ to the joint exponent.

Now, for sufficiently small $s_a$, the qualitative behaviour of the flow does change in the region with $|\phi|>\pi$. This only changes our discussion regarding the further deformation along the flow of the $r_h>s_a/3$ portions, because what happens then is that the thimble associated to the joint exponent saddle at $\phi=0$ does not end up in a saddle at $\phi=-2\pi\equiv2\pi$, but actually asymptotes to the line $\phi=\pm2\pi$. However, in this situation we again have a further deformed contour in which the integral is ``manifestly convergent and saddle point dominated'', because it is still made of steepest descent flow lines.

Hitherto these are just fixed-length saddles, there is no guarantee to have extrema with respect to $s_a$. However, as noted, there is actually a maximum when $s_a=s_a^*$ and one has $W_\text{HJ}(s_a^*)\approx1.85\times S_{\text{dS}_3}$, where $S_{\text{dS}_3}=
\frac{4\pi^2}{\sqrt\Lambda}$ denotes the Gibbons-Hawking entropy.
In a semiclassical expansion this maximum would dominate, and therefore one would have
\begin{equation}
    Z\sim \exp\left(1.85\times S_{\text{dS}_3}\right).
    \label{eq:Z_semi_classical}
\end{equation}
Due to its Euclidean nature and the fact that its action scales with $S_{\text{dS}_3}$ as $1/{\sqrt\Lambda}$, this is identified as a discrete Euclidean de Sitter saddle.
The result \eqref{eq:Z_semi_classical} was reached with a Lorentzian path integral and resembles that of Gibbons and Hawking, up to the numerical factor 1.85 (which we will discuss in \S\ref{ssec:refinement}). However, before we can conclude that the Euclidean saddle dominates, we must evaluate the contribution coming from the $s_a^\Lambda<s_a<\infty$ domain.

The exponential enhancement, which resulted from the imaginary contribution due to the chosen side of the branch cut, appears at first to lead to trouble. The remaining contribution to the path integral, from the $s_a\in(s_a^\Lambda,\infty)$ regime, also contains the branch cut portions of the contour, and the corresponding exponential enhancement grows indefinitely with the size of the horizon, \textit{i.e.} $s_a$ itself. Note, however, that we integrate over \textit{both} positive and negative lapse. The net integral depends on how these contributions combine. In the Euclidean saddle regime, 
as explained above, the contour can be deformed away from the branch cut to pass through a saddle that is also exponentially enhanced and dominates the integral. Thus, in the Euclidean saddle regime, the contributions from the two portions are evidently additive. In the Lorentzian saddle regime, on the other hand, it turns out that they cancel.\footnote{One can in fact show that they do not vanish individually.}

To show that they cancel we start by noting that Cauchy's theorem ensures, that the fixed-length integral equals the integral of the arc at infinite radius going from $\phi=-\pi$ to $\phi=\pi$. More precisely for the fixed-length integral in the regime $s_a>s_a^\Lambda$, 
we have
\begin{align}
    Z_{s_a>s_a^\Lambda}&=\lim_{r_h\rightarrow\infty}\int_\text{arc}\mathrm d s_h\,\mu_h \,e^W\nonumber\\
    &=\lim_{r_h\rightarrow\infty}\int_{-\pi}^\pi\mathrm d\phi\, (\imath r_h e^{\imath\phi})\mu_h\, e^W.
    \label{eq:fixed_length_integral}
\end{align}
We will prove that this infinite arc integral is zero provided the measure 
satisfies $|\mu_h|= r_h^{\alpha}$ with $\alpha < -1/2$.
In particular, this is the case for our measure \eqref{eq:measure}.

The key intuition behind this is based on the fact that the asymptotic behaviour of the exponent changes as one crosses the threshold $s^\Lambda_a$ from below: As seen from eq. \eqref{eq:W_asymptotics} and suggested by the gradient flow of $\Re  W$ in FIG.~\ref{fig:flows}, in the region $-\pi<\phi<\pi$ the limit of $\Re  W$ as $r_h\rightarrow\infty$ changes from diverging to $\infty$ (for $s_a<s_a^\Lambda$) to diverging to $-\infty$ (for $s_a>s_a^\Lambda)$. This is a manifestation of the  competition between the curvature term and the cosmological constant term in the action. Therefore, for large $r_h$, the integrand in \eqref{eq:fixed_length_integral} is exponentially suppressed. However, $\Re W(\pm\pi)= 0$ for $r_h>s_a/3$, so right at the boundary points $\phi=\pm \pi$ of the arc the exponential suppression disappears and
the absolute value of the integrand is of order $r_h \mu_h$.
This could lead to a 
nonzero and even divergent result
if $\mu_h$ does not (sufficiently) suppress large heights. We shall now see that with a suitable measure the integral indeed vanishes.\footnote{There is an alternative approach that would lead to the same conclusion, while using a trivial measure $\mu_h=1$. One can regularize the purely oscillatory portions of the fixed-length integrals (\textit{i.e.} those with $s_a/3<r_h<\infty$) in the standard way by adding an exponential damping for large $r_h$ parametrized by an infinitesimal parameter $\epsilon$. This can be achieved by deforming the contours infinitesimally away from the Lorentzian lines for large heights. However, given how the asymptotic behaviour of $\Re W$ changes when crossing $s_a^\Lambda$, the deformation would depend on the regime in question. For $s_a>s_a^\Lambda$ the contours would be positioned asymptotically at $\phi=\pm\pi\mp\epsilon$ and the integral over the asymptotic arc would be zero, because the arc now does not include the boundary points at $\phi=\pm\pi$. For $s_a<s_a^\Lambda$ one would have a contour asymptotically at $\phi=\pm\pi\pm\epsilon$ and the same Euclidean saddle point as in the main text dominates in a semiclassical expansion. However, this asymptotic deformation for $s_a<s_a^\Lambda$ violates the off-shell (strong) version of the convergence criterion. Thus, although the main conclusions of \S\ref{ssec:saddle_analysis} would hold also in this approach, it comes with the cost of letting go of the strong convergence criterion and adopting instead its weak (on-shell) version, \textit{i.e.}, that in which it is only required for the dominating saddle point to be such that $\Re N>0$. Further work is needed to determine whether 
the weak version suffices in the context of the full multidimensional integral.
}

\begin{figure}
    \centering
    \includegraphics[width=0.45\textwidth]{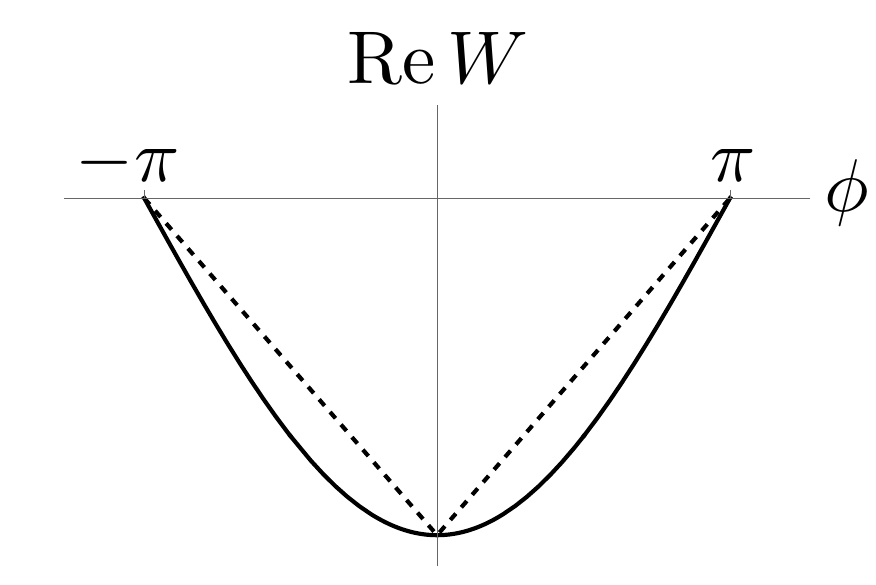}
    \caption{
    This figure shows $\text{Re}\,W$ (solid) as a function of the Wick rotation angle $\phi$ in the range $\phi \in [-\pi,\pi]$, with $s_a \Lambda=50$ and $r_h\Lambda=10^3$. For sufficiently large $r_h$, $\text{Re}\,W$ is non-positive and convex for $\phi \in [-\pi,\pi]$. It can therefore be bounded by a piecewise linear function (dashed) obtained by gluing secants. Here we choose to use two secants, which cut the $\text{Re}W$ graph at $\phi=-\pi$ and $\phi=0$, and at $\phi=0$ and $\phi=+\pi$, respectively.
    }
    \label{fig:bound}
\end{figure}

We begin by observing that
for sufficiently large $r_h$, $\Re W$ is convex as a function of $\phi \in [-\pi,\pi]$, as illustrated by the solid line in FIG.~\ref{fig:bound}. Therefore its graph lies below the two secants which go from the point $(-\pi,0)$ to $(0,\Re W(\phi=0))$ and from the point $(0,\Re W(\phi=0))$ to $(\pi,0)$, respectively (\textit{cf.} dashed lines in FIG.~\ref{fig:bound}). In other words, $\Re W$ is bounded from above by the function
\begin{equation}
    \beta:={\Re W(\phi=0)}(1-|\phi|/\pi),
    \label{eq:bound}
\end{equation}
whose graph is illustrated by the dashed line in FIG.~\ref{fig:bound}. Thus, 
\ba
0 \le  |Z_{s_a>s_a^\Lambda}|
&\le&
\lim_{r_h\rightarrow\infty}\int_\text{arc}\left|\mathrm d s_h\mu_h e^W\right|\nn\\
&\le&
    \lim_{r_h\rightarrow\infty}\int_{-\pi}^\pi\mathrm d \phi\, r_h^{1+\alpha}\, e^\beta \nn\\
    &=&
    \lim_{r_h\rightarrow\infty} -\frac{2\,\pi r_h^{1+\alpha}}{W(\phi=0)}
    \nn\\
   & \propto &
   \lim_{r_h\rightarrow\infty} r_h^{\frac12 + \alpha}\,\,
    \label{eq:integral_triangle_inequality}
\ea
where we have used \eqref{eq:measure}.
The last two relations follow because (\textit{cf.} \eqref{eq:W_asymptotics})
$$\lim_{r_h\rightarrow\infty} W(\phi=0)\propto \lim_{r_h\rightarrow\infty}  (-\sqrt{r_h}) =-\infty.$$
The upper bound thus vanishes provided $\alpha < -1/2$.

In summary, above the threshold, the fixed-length integral over the arc at infinity connecting the Lorentzian branches is zero, as its absolute value is bounded from above by the integral of $ r_h^{1 + \alpha} |e^\beta|$, which vanishes if $\alpha<-1/2$. But since said arc together with the original integration domain form a closed contour, it must be that the integration over the latter also vanishes, by virtue of Cauchy's theorem. 
(Note also that we can {\it not} conclude the same in the $(0,s_a^\Lambda)$ sub-domain, since according to \eqref{eq:W_asymptotics} the integrand diverges for $r_h\rightarrow\infty$.)

As said cancellation happens for every $s_a$ with $s_a^\Lambda<s_a<\infty$, this second regime of the partition function path integral does not contribute, and therefore in the semiclassical approximation we indeed recover, 
by deformation of a Lorentzian contour,
the Gibbons-Hawking-like result \eqref{eq:Z_semi_classical} as conjectured in the continuum setting of \cite{Banihashemi:2022jys,Jacobson:2022jir}. 

We note that there is also a more heuristic reason why the saddle points along the Lorentzian lines do not contribute: these are only saddle points if we restrict variations to the $s_h$ variable, but none of these partial saddle points turns out to be a saddle with respect to variations of both $s_h$ and $s_a$.

Thus, using our discrete formulation we have found that,
starting from a quasi-Lorentzian path integral 
for the partition function under study, 
a) one obtains an exponentially enhanced result for the entropy consistent with the result of Gibbons and Hawking, which arises due to the imaginary contribution to the action that comes from CTC singularities, and b) although the exponential enhancement 
of the integrand grows without bound with the size of the system boundary,  
its contribution to the integral is cut off by cancellations, via a mechanism
similar to that discussed in a related continuum context in ref.~\cite{Marolf:2022ybi}.

Note, however, that the overall factor in the exponent of \eqref{eq:Z_semi_classical} makes it such that our result does not exactly match that of Gibbons and Hawking, which agrees with the Bekenstein-Hawking entropy formula for a cosmological horizon
of circumference (``area'') $L={2\pi}/{\sqrt\Lambda}$, namely
$S_{\text{dS}_3}=2\pi L={4\pi^2}/{\sqrt{\Lambda}}$.\footnote{Recall that in our units $8\pi G=\hbar=1$, so the Bekenstein-Hawking entropy for a horizon is $L/4\hbar G = 2\pi L$.} 
Thus, the precise continuum dS result is not recovered. This is likely a discretization artifact. Indeed, the three-dimensional Regge Hamilton-Jacobi function is not triangulation invariant in the presence of a non-zero cosmological constant when using flat\footnote{The situation changes when dealing with homogeneously curved tetrahedra \cite{Improved}.} tetrahedra \cite{Diff1}, as done here; thus one would expect to recover the continuum result only in a continuum limit.

Before commenting further on this issue of discretization artifacts 
let us contemplate what would happen if we were to choose the Lorentzian contour to go along the exponentially suppressing side of the branch cut. In this case both the contour part corresponding for positive lapse and the contour part corresponding to negative lapse can be deformed so as to pass through the Euclidean saddle point at $\phi=2\pi$, which give exponentially suppressed contributions (\textit{cf.} \cite{Asante:2021phx}). In case one is using a measure $\mu_h\sim s_h^{-1}$, which does allow us to use the Riemann surface coordinatized by $\phi \in [-2\pi,2\pi]$,  one can even close the contour with an arc at infinity, whose contribution is zero. One would thus find that the integral over $s_h$ leads to zero in the Euclidean saddle regime.
The appearance of the branch cuts and the choice of branch cut side seem therefore to be essential for the result.

Notably, our branch cuts may have a continuum analogue: As discussed above, our calculation is structurally similar to continuum Lorentzian minisuperspace no-boundary wave function calculations which can be reduced to an integral over a global lapase that is comparable with our fixed-length integrals. The continuum action includes however a $1/N$-term (originating from 
proper time derivatives normal to the spatial hypersurfaces), hence the $N=0$ point represents an essential singularity. By contrast, the Regge action is finite for vanishing height, but has a branch cut for sufficiently small height values. The branch cut in the discrete theory and the $1/N$ singularity in the continuum both arise from degeneracy of the metric. When $N$ goes to zero, the spacelike foliation degenerates, and no proper time flows. In our simplicial setting, this happens at the CTC singularity even when the timelike height of the tetrahedron is nonzero, provided it is small enough that there is a CTC singularity where time does not flow.

\begin{figure}
    \centering
    \includegraphics[width=0.48\textwidth]{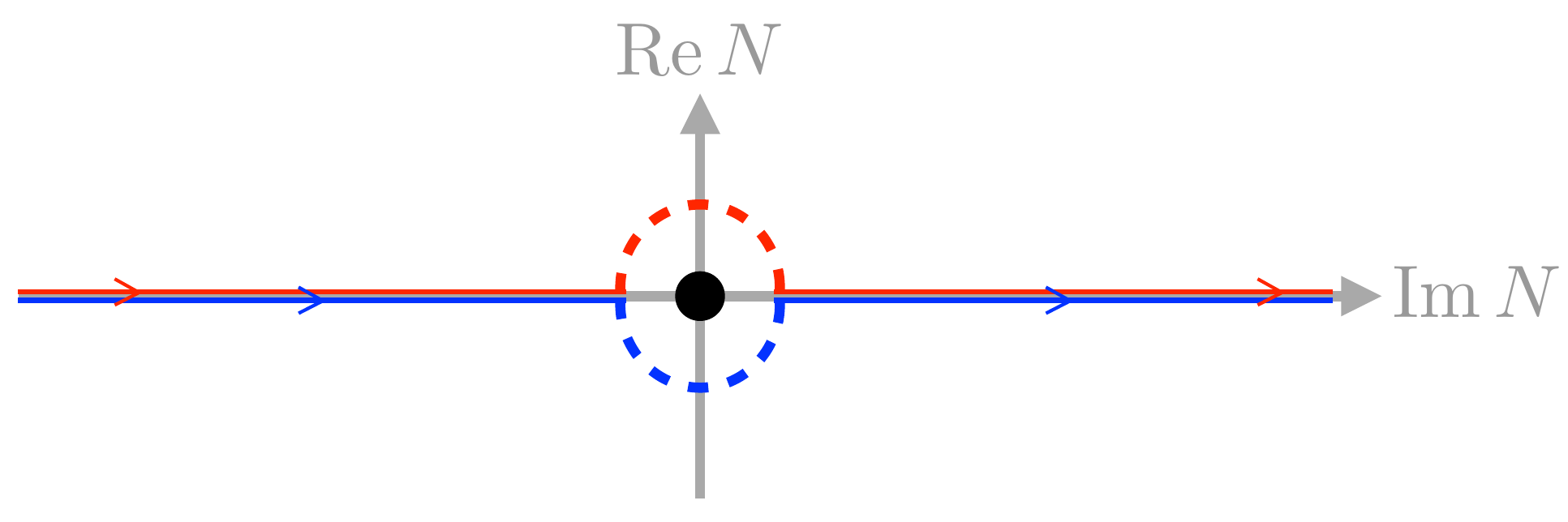}
    \caption{Different global lapse contours (red/upper and blue/lower dashed lines) used in the literature to compute the no-boundary wave function from path integrals \cite{Feldbrugge:2017fcc,DiazDorronsoro:2017hti}. They differ in how they circumvent an essential singularity (black dot) at $N=0$. This continuum situation is analogous to ours, where we may navigate the branch cuts, which are centered at $s_h=0$, in two different ways if we do not take into account the convergence criterion. This similarity suggests that the Regge action may in general blow up singularities of the continuum action as branch cuts.}
    \label{fig:continuum_contours}
\end{figure}

Whereas in the discrete case one needs to decide how to navigate the branch cut for small height values, one needs to decide in the continuum how to navigate the essential singularity at $N=0$. One can indeed reproduce both the choice of the suppressing side of branch cut and the enhancing side of branch cut in the continuum mini-superspace discussion, as shown by the works \cite{Feldbrugge:2017fcc} and \cite{DiazDorronsoro:2017hti}, respectively. These works indeed differ in their choice of how to navigate the essential singularity. The different contour choices are illustrated in FIG.~\ref{fig:continuum_contours}.  In fact, when choosing the side leading to exponential enhancement one picks up a contribution from the Hartle-Hawking saddle, so that after setting $a_1=0$ (which corresponds to integrating over all minisuperspace geometries with no-boundary at all, similar to our calculation) one gets $Z\sim\exp S_\text{dS}$ in a semiclassical limit, as do we.

One could thus say that by discretization we have blown up the essential singularity into a branch cut. This has the advantage of making the related choice of contour more obvious and also provides an intuitive explanation of how Lorentzian path integrals can lead to exponentially enhanced results. Further, it could be interesting to explore if the essential singularity and corresponding circumvention in the lapse complex plane somehow capture the geometry of CTC singularities or a remnant of it. Indeed, we 
just argued  that the continuum mini-superspace path integral (with the appropriate choice of contour near $N=0$) leads also to the de Sitter entropy. This happens despite the fact that the continuum mini-superspace geometries do not include CTC singularities---at least as long as the scale factor and lapse are kept real. The match between the discrete and continuum mini-superspace results suggests, however, that the geometries resulting from making the lapse complex might describe CTC singularities.

\subsection{Refinement of the discretization \label{ssec:refinement}}

We noted the numerical disagreement between the continuum result $S_{\text{dS}_3}={4\pi^2}/{\sqrt{\Lambda}}$ and the discrete one coming from \eqref{eq:Z_semi_classical}. To evaluate whether it dissipates in the continuum limit we can check if the discrepancy reduces when refining the triangulation. Let us do so.

We will consider a minimal refinement scheme that will indeed reduce the difference between the continuum and discrete results. However, it will not be a continuum limit. In particular, we will not be adding more degrees of freedom into the system, but we will make it smoother. Thus, the difference is not eliminated completely, but enough to support the expectation that the discrepancy is a discretization artifact.

\begin{figure}[!h]
    \centering
    \includegraphics[width=0.48\textwidth]{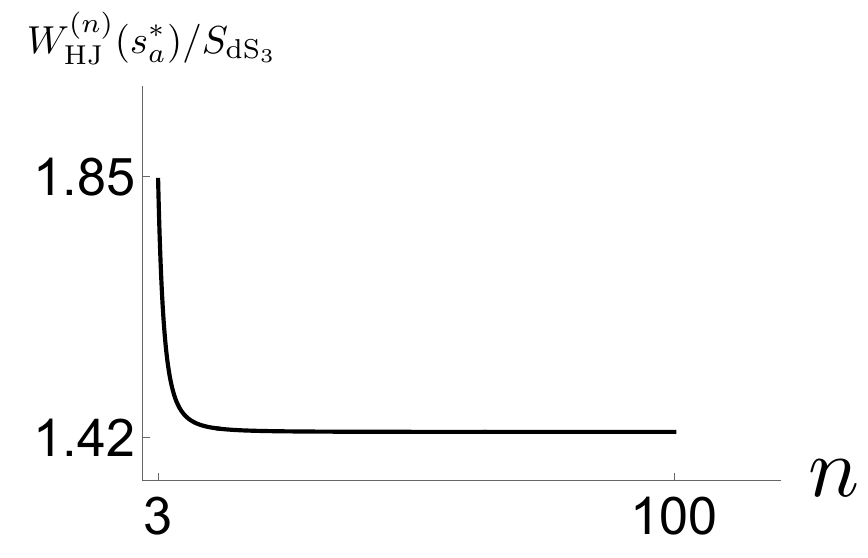}
    \caption{Fitted function showing the ratio of the continuum and discrete computations. The ratio goes closer to one with the refinement scheme, but saturates around $1.42$}
    \label{fig:refinement}
\end{figure}

So far we used as a fundamental building block an isosceles triangular pyramid with equilateral base triangle. This can be refined (in the sense of smoothing the horizon's triangulation) by an isosceles pyramid with square base, or pentagonal base, or in general a \textit{regular} $n$-gon as a base. We can then glue four of these pyramids as in FIG.~\ref{fig:3D_discrete} to obtain a sequence of triangulations labeled by $n$. In the $n\rightarrow\infty$ limit the sequence reproduces the geometry of four cones glued together as in the figure.
The analytic structure of any of these triangulations is qualitatively similar to that of the original tetrahedral case, so the arguments of \S\ref{ssec:saddle_analysis} migrate directly and we can ask what is the semiclassical behaviour of the partition function as $n$ changes. This is captured in FIG.~\ref{fig:refinement}, which confirms the discussion above, namely: as we refine, the continuum result is approached, but not reached, as this is not the full continuum limit. This suggests that, indeed, the pre-factor in the exponent of \eqref{eq:Z_semi_classical} is a discretization artifact.
\section{Discussion \label{sec:discussion}}

 The Lorentzian gravitational path integral (integrating over both positive and negative lapse) imposes the diffeomorphism and Hamiltonian constraints \cite{Halliwell:1988wc} onto the initial and final states and computes the physical inner product between states. 
 If the initial and final configurations are identified,  the path integral 
 computes the dimension of the physical Hilbert space. 
In this work we implemented this in the case of simplicial minisuperspace
gravity in 2+1 dimensions with a postive cosmological constant, 
for a system whose spatial configuration is
a triangle, and whose spacetime topology is a 3-sphere. The configurations
are characterized by two squared edge lengths, and
include a class with closed timelike curves around the perimeter of the triangle, 
as required if the path integral is to compute the dimension of the 
Hilbert space of triangle geometries. We expected the result 
to agree, at least qualitatively, with the 
entropy associated with a static patch of de Sitter space,
and this expectation was confirmed. 
Key to the result is the imaginary contribution to the action that arises
from the presence of the closed timelike curves, in close analogy with
what was found in a related continuum calculation of black hole entropy~\cite{Marolf:2022ybi}.

The simplicial framework offers a number of advantages and insights:

\begin{itemize}

\item
It provides a regularization of the path integral, turning it into a finite dimensional integral. This allows a much more explicit treatment of the path integral than in the infinite dimensional continuum case.  
The two-dimensional integral naturally splits into an integral where we keep the area (here a length) of the support of the conical singularity fixed, and a second integration over this volume. 
The same split has been discussed for the continuum in \cite{Marolf:2022ybi}.

\item The Lorentzian Regge action, derived via analytic continuation from the Euclidean Regge action \cite{Asante:2021phx}, naturally includes imaginary terms for codimension-2 conical singularities which support light cone irregularities \cite{Sorkin:2019llw,Asante:2021phx}. This is to be contrasted with the continuum framework \cite{Colin-Ellerin:2020mva,Marolf:2022ybi}, where additional input is needed in order to define the action in the presence of such light cone irregularities. 

\item The framework of the complex Regge action \cite{Asante:2021phx},  which we employ here, makes also clear that (edge) light cone irregularities are associated with branch cuts. The Lorentzian contour is therefore not uniquely defined in the presence of such light cone irregularities. One rather needs also to specify how to navigate the branch points. In doing so the contour cannot remain strictly Lorentzian, as we have to go into the complex domain. (There is a second sense in which our contour is not strictly Lorentzian, as we do allow light cone irregularities to occur.)  

The choice of contour along the branch cut decides whether the resulting path integral picks up exponentially enhancing or suppressing contributions. These result from the imaginary terms in the Regge action, and explain how a Lorentzian path integral can lead to an exponentially enhanced result and thus a large entropy.

The exponential enhancement is present already along our quasi-Lorentzian integration contour, but we are also able to rigorously justify deformation of the contour so that it passes through a Euclidean saddle that yields the leading order result and is a discrete version of the one postulated by Gibbons and Hawking \cite{GibbonsHawking} to be dominant. Earlier work \cite{Banihashemi:2022jys} argued that such a contour deformation should be possible.  Here, in the simplicial minisuperspace setting, we have verified that indeed it is.

One might think that the ambiguities in the Lorentzian contour are a discretization artifact, however this does not seem to be the case. A related choice of contour exists also in continuum calculations of the no-boundary wave function, hidden in the choice of detour around an essential singularity of the integrand at vanishing lapse \cite{Feldbrugge:2017kzv,DiazDorronsoro:2017hti,Marolf:1996gb}. The same kind of ambiguity appears also in the continuum discussion of topology changing trouser configurations \cite{Louko:1995jw}.

\item  Our discrete example offers sufficient control over the integral 
so that we can establish explicitly its convergence along the contour dictated by the 
nature of the partition function being calculated, 
namely, the quasi-Lorentzian contour described above.
We are also able to establish
that a contour deformation to certain Lefschetz thimbles is allowed.\footnote{See \cite{Dittrich:2023rcr} for a numerical technique (known as acceleration of series convergence) which allows to evaluate the Lorentzian integral without any contour deformation. The results coincide with a numerical integration along a deformed (Lefschetz thimble) contour. } 
Unlike in some work on the path integral for a cosmological 
wave function (\textit{e.g.}, \cite{Lehners:2021mah}), the thimble contour is 
here justified from first principles via deformation of the original
partition function contour, rather than being postulated.

\item In addition to the convergence of the minisuperspace integral taken by itself, 
there is the crucial issue of convergence of the (infinitely many) 
``fluctuation'' integrals that are ignored in the minisuperspace treatement. 
As first emphasized by Halliwell and Hartle \cite{HalliwellContours} in the context
of the saddle point approximation to the 
no-boundary wave function, the recovery of quantum field theory in curved spacetime
requires that the fluctuation integrals are convergent, which in turn requires that 
the real part of the lapse function be positive (in a convention where the Euclidean lapse is real). (See also \cite{Louko:1995jw,Witten:2021nzp,Kontsevich:2021dmb} for 
related discussions.)

We have adopted this fluctuation convergence criterion (FCC) to determine
how to circumnavigate the branch points of the Lorentzian contour, and found that
the resulting contour is also the one on which there is exponential enhancement,
rather than exponential suppression, of the integral. 
It makes good sense that satisfaction of the FCC is required in order
to obtain the large Bekenstein-Hawking horizon entropy, given that this
 entropy can be understood as a form of vacuum entanglement entropy, 
UV regulated somehow by quantum gravity \cite{Jacobson:2012yt}:
the FCC is essentially the condition that 
the fluctuations behave locally as they do in the Minkowski vacuum.

\item As emphasized in \cite{Marolf:2022ybi} the mechanism allowing for a positive entropy carries however also the danger of giving a divergent result: we also need to integrate over the area (here given by a length parameter $s_a$) of the conical singularities. With exponential enhancement this integral seems bound to diverge. However, we encounter a novel and surprising mechanism that prevents 
such a divergence. The threshold value $s_a^\Lambda=8\sqrt{3}\pi/\Lambda$ 
distinguishes two regimes. For $s_a<s_a^\Lambda$ the integral (over the remaining parameter, which can be identified with a lapse) has only Euclidean saddles, whereas for $s_a>s_a^\Lambda$ the system has two Lorentzian saddles, one for positive and one for negative lapse. Integration over both positive and negative lapse---as required for projection onto the physical Hilbert space---leads to cancellation of these large $s_a$ contributions; 
in fact, a closed contour argument shows that 
in this regime the integral vanishes exactly.

\end{itemize}

Also connected with these fixed $s_a$ integrals over our lapse-like variable is the observation we made in \S\ref{ssec:saddle_analysis} pointing out that they are structurally similar to real time no-boundary wave function computations in the Regge framework \cite{Hartle1,Hartle2,Hartle3,Asante:2021phx}. It is therefore tempting to compare the analytic structure with the one of the corresponding continuum minisuperspace calculations where one sees that at small lapse one needs to leave the strictly Lorentzian contour in order to circumvent an essential singularity at zero lapse. This seems to be reminiscent  of the branch cut circumvention in the discrete needed for small lapse, so it would be interesting to study this similarity in greater detail.

It is also interesting to compare the behavior of the fixed $s_a$ Euclidean saddle action with that in the analogous continuum setting. As seen in FIG.\ \ref{fig:fixed_length_HJ},  the negative of the simplicial saddle action grows with $s_a$ to a maximum at some $s_a^*$, and then decreases to zero at $s_a^\Lambda$. Moreover, the lapse ($\sqrt{r_h}$)  at the saddle diverges in the
limit $s_a\rightarrow s_a^\Lambda$. In the continuum, the minimal action Euclidean saddle at fixed spatial volume was found in \cite{Jacobson:2022jir} to behave similarly in both of these respects. A difference, however, is that in the simplicial case the boundary perimeter of the spatial triangle reaches the maximum value $3\sqrt{s_a^\Lambda}>0$ when the action vanishes, whereas in the continuum case the action is proportional to the boundary $(D-2)$-area, which goes to zero when the spatial $(D-1)$-volume covers the complete de Sitter equatorial $(D-1)$-sphere.

For the case of zero cosmological constant,
$s_a^\Lambda$ goes to infinity,  so a Euclidean saddle exists
for any fixed spatial area. In the simplicial model, at the saddle
in this limit we have $r_h\sim s_a^2\Lambda\rightarrow0$.
One can see by inspection of the general formula for $W$ in \eqref{eq:W_a} that in this limit $W\rightarrow 6\pi\sqrt{s_a}$, which remarkably
is precisely equal (in our units with $8\pi G\hbar = 1$)
to the Bekenstein-Hawking entropy associated with 
the triangle perimeter, $L=3\sqrt{s_a}$. 

Another comparison between the simplicial and continuum cases worth mentioning concerns the curvature of the fixed spatial area Euclidean saddle geometry. In the continuum, unless the fixed spatial area corresponds precisely to that of the de Sitter hemisphere, there is a mild (integrable) curvature singularity at the disc 
boundary (\textit{i.e.}, at the Euclidean horizon), which implies that higher curvature terms that would appear in the effective field theory action for a UV completion of the theory should be taken into account. In the simplicial theory, on the other hand, all curvature is focused onto bones, and the discrete analog of curvature never diverges.  

We have seen that the simplicial Regge framework can shed light on a number of issues for the Lorentzian gravitational path integral. Numerous generalizations are possible, and promise to yield interesting physical insights. A simple variation on what we have done here
would be to employ a triangulation whose configurations, for any value of the ``lapse'', are only those 
with CTC singularities, \textit{i.e.}, which do not include initial and final singularities. This would be closer to the nature of the continuum partition function. And it would be interesting to devise a continuum
minisuperspace partition function that would involve integration over such configurations, which could be compared with the simplicial model.
More generally, refining our triangulation and relaxing symmetry assumptions the Regge framework would allow one to add geometric fluctuations, such as anisotropies, in a controlled manner, in analogy with the continuum minisuperspace work in \cite{Feldbrugge:2017mbc, 
DiazDorronsoro:2018wro,
Lehners:2024kus}. This would clarify the role of the fluctuation convergence criterion, and more generally could give us more confidence that the Lorentzian gravitational path integral can be made well defined.
And, finally, these methods could be applied to
investigations of black hole space times and configurations which describe topology change.

\begin{acknowledgments}
 
We thank Seth Asante for providing FIGURE~\ref{fig:Lorentz_angles}. JPA thanks Oleksandra Hrytseniak and Donald Marolf for discussions and is supported by an NSERC grant awarded to BD.  Part of this research was conducted while BD was visiting the Okinawa Institute for Science and Technology (OIST) through the Theoretical Sciences Visiting Program (TSVP). Research at Perimeter Institute is supported in part by the Government of Canada through the Department of Innovation, Science and Economic Development Canada and by the Province of Ontario through the Ministry of Colleges and Universities.
TJ is grateful to PI for a hosted research visit during which this collaboration was initiated.
The work of TJ was supported also in part by 
NSF grants PHY-2012139 and PHY-2309634. 

\end{acknowledgments}

\appendix   

\section{Convergence of the $s_h$ integral along the Lorentzian contour
}\label{App:Dirichlet}

Here we  establish the (conditional) convergence of the fixed length path integral $Z_{s_a}=\int dr_h \mu(r_h)e^W(r_h)$
along the Lorentzian contour when  $r_h\rightarrow \infty$. To do so we apply Dirichlet's test for convergence of improper integrals \cite{Shilov}. This states that the integral
\ba
\int_a^\infty f(x) g(x) \text{d} x
\ea
of the product of two continuous complex valued functions converges if 
two conditions hold:
$(i)$ the 
modulus of $\int_a^x f(t) \text{d}t$ is uniformally bounded on all intervals in $[a,\infty)$, and $(ii)$ the real and imaginary parts of $g$ are monotonic and
$\lim_{x\rightarrow \infty}g(x)=0$. To apply this test we  make use of the asymptotic expansion of $W(r_h)$ for large $r_h$. According to equation (\ref{eq:W_asymptotics}) we have for $\phi=\pm \pi$ the asymptotic behaviour
$W(r_h)\sim \imath c \sqrt{r}$, with $c\in \mathbb{R}$. The subleading terms are of the form  $\imath R(r_h)=\imath \sum_{k\geq 1}c_k r_h^{-k+1/2}$ with $c_k\in \mathbb{R}$,  hence $R\rightarrow 0$ for $r_h\rightarrow \infty$. For large $r_h$ the integral thus becomes
\begin{gather}
    \int_a^\infty dr_h \,\, r_h^\alpha \exp(\imath R(r_h))\exp(\imath c \sqrt{r_h})\nonumber\\
    =\nonumber\\
    2 \int_{\sqrt{a}}^\infty \text{d} x\,\,x^{2\alpha+1} \exp(\imath R(x^2))  \exp(\imath c x),
\end{gather}
where $x\equiv\sqrt{r}$. We then choose $f=\exp(\imath x)$ and $g=x^{2\alpha+1} \exp(\imath R(x^2))$. Dirichlet's test is passed as long as $\alpha<-1/2$ and we choose the lower boundary value $a$ for the integral sufficiently large, so that the imaginary and real parts of $g(x)$ are monotonic (remember that $R(x)\rightarrow 0$ for $r\rightarrow 0$). In the main text we show that $\alpha<-1/2$ also ensures that the integral over a certain $\phi$-parametrized arc  vanishes in the limit of $r_h\rightarrow \infty$, which allows us to replace the original contour with a deformed contour.

\section{Reality of the partition function \label{App:reality}}

Here we show that the simplicial path integral computed in the main text, $Z$, is real, which is consistent with it being a discrete version of a partition function computing the dimension of a Hilbert space.

To do so we begin by splitting $Z$ (\textit{cf.}\ \eqref{eq:Z_path_integral}) into its contribution from the left and right quasi-Lorentzian portions of the contour at $\phi=\mp\pi$, respectively, together from the contribution of the arc around $r_h=0$ (\textit{cf.}\ FIG.~\ref{fig:flows}):
\begin{align}
    Z&=\lim_{\epsilon\rightarrow0}\int_0^\infty\mathrm ds_a\mu_a(s_a)Z_{s_a}\nonumber\\
    &=\lim_{\epsilon\rightarrow0}\int_0^\infty\mathrm ds_a\mu_a(s_a)\left(Z_{s_a}^\text{L}+Z_{s_a}^\text{R}+Z_{s_a}^\text{arc}\right).
    \label{eq:Z_reality_split}
\end{align}
As stated in \S\ref{ssec:Regge_action} we consider a real $\mu_a$, so it suffices to show that the fixed length-path integrals $Z_{s_a}$ with measure \eqref{eq:measure}
are real. We begin with $Z_{s_a}^\text{arc}$.
\begin{align}
    Z_{s_a}^\text{arc}&=\int_\text{arc}\mathrm ds_h\mu_h e^W=-\imath\int_{-\pi}^\pi\mathrm d(\epsilon e^{\imath\phi})(\epsilon^\alpha e^{\imath\alpha\phi})e^W\nonumber\\
    &=\epsilon^{1+\alpha}\left(\int_{-\pi}^0\mathrm d\phi \, e^{\imath(1+\alpha)\phi}e^W+\int_0^\pi\mathrm d\phi \, e^{\imath(1+\alpha)\phi}e^W\right),
\end{align}
but
\begin{equation}
    \int_{-\pi}^0\mathrm d\phi \, e^{\imath(1+\alpha)\phi}e^{W(\phi)}=\int_0^\pi\mathrm d\phi \, e^{-\imath(1+\alpha)\phi} e^{W(-\phi)},
\end{equation}
and therefore using the fact that (see \cite{Asante:2021phx} for a proof)
\begin{equation}
    W(\phi)=W(-\phi)^*
    \label{eq:W_reflective_property}
\end{equation}
we have
\begin{equation}
    Z_{s_a}^\text{arc}=\epsilon^{1+\alpha}\left(\int_0^\pi e^{\imath(1+\alpha)\phi}e^W\mathrm d\phi+\text{c.c.}\right),
\end{equation}
which is manifestly real.

Similarly, $Z_{s_a}^\text{L}+Z_{s_a}^\text{R}$ is real, because these terms are complex conjugates of each other, as the following argument shows.
Let us denote $\mathcal C_\text{L}$ as the left $\epsilon$-deformed contour and similarly for the right one, and their parametrization as $s_h^\Box(r_h)=r_h e^{\imath\phi^\Box(r_h)}$. From FIG.~\ref{fig:flows} it is clear that
\begin{equation}
    \phi^\text{L}(r_h)=-\phi^\text{R}(r_h)\quad\Rightarrow\quad s_h^\text{L}(r_h)=s_h^\text{R}(r_h)^*,
\end{equation}
so using \eqref{eq:W_reflective_property} we have
\ba
 Z_{s_a}^\text{L}&=&-\imath\int_{\mathcal C_\text{L}}\mathrm ds_h s_h^\alpha e^W=\imath\left(\int_{\mathcal C_\text{R}}\mathrm ds_h s_h^\alpha e^W\right)^*
 \nn\\
 &=&
 \imath\left(\imath Z_{s_a}^\text{R}\right)^*
 =\left(Z_{s_a}^\text{R}\right)^*,
    \label{eq:branches_are_conjugate}
\ea
where the sign change in the second equality is due to the fact that $\mathcal C_\text{L}$ and $\mathcal C_\text{R}$ are traversed in opposite directions.
From \eqref{eq:branches_are_conjugate} if follows that $Z_{s_a}^\text{L}+Z_{s_a}^\text{R}$ is indeed real, and because so is $Z_{s_a}^\text{arc}$, it follows from \eqref{eq:Z_reality_split} and the reality of $\mu_a$ that $Z$ is real.

\bibliography{references}

\end{document}